\documentstyle[12pt,fleqn,epsfig,amssymb]{article}
\newcommand {\bR}{{\Bbb R}}
\newcommand {\bN}{{\Bbb N}}
\newcommand {\bZ}{{\Bbb Z}}
\newcommand {\bC}{{\Bbb C}}

\newcommand {\bT}{{\Bbb T}}
\newcommand {\bP}{{\Bbb P}}

\newcommand {\cC}{{\cal C}}
\newcommand {\cD}{{\cal D}}
\newcommand {\cF}{{\cal F}}
\newcommand {\cI}{{\cal I}}
\newcommand {\cG}{{\cal G}}
\newcommand {\cL}{{\cal L}}
\newcommand {\cO}{{\cal O}}
\newcommand {\cS}{{\cal S}}
\newcommand{\beq}{\begin{equation}}
\newcommand{\eeq}{\end{equation}}
\newcommand{\beqn}{\begin{eqnarray}}
\newcommand{\eeqn}{\end{eqnarray}}
\newcommand{\beqno}{\begin{eqnarray*}}
\newcommand{\eeqno}{\end{eqnarray*}}
\newtheorem{theorem}{Theorem} [section]
\newtheorem{lemma}[theorem]{Lemma}
\newtheorem{propo}[theorem]{Proposition}
\newtheorem{defi}[theorem]{Definition}
\newtheorem{coro}[theorem]{Corollary}

\newtheorem {remark}[theorem]{Remark}
\newtheorem {remarks}[theorem]{Remarks}
\newtheorem {conject}[theorem]{Conjecture}

\renewcommand {\l}{\left}
\newcommand {\r}{\right}
\newcommand {\q}{q} 
\newcommand {\p}{p} 
\newcommand {\LA}{\left\langle}
\newcommand {\RA}{\right\rangle}
\newcommand {\vep}{\varepsilon}

\newcommand {\om}{\omega}
\newcommand {\ar}{\rightarrow}

\newcommand {\eh}{\textstyle \frac{1}{2}}
\newcommand {\ph}{\varphi}
\newcommand {\SE}{{\Sigma_E}}
\newcommand {\Vmax}{{V_{\rm max}}}
\newcommand {\Vmin}{{V_{\rm min}}}
\newcommand {\Vmean}{{V_{\rm mean}}}
\newcommand {\gEh}{{\hat{g}_E}}
\newcommand{\OO}[1]{{{\cal O}\left(\hbar^{#1}\right)}}
\newcommand {\KE}{K_{E}}
\newcommand {\Eth}{{E_{\rm th}}}
\newcommand {\hP}{{\hat{P}}}
\newcommand {\hH}{\hat{H}}
\newcommand {\hp}{\hat{p}}
\newcommand {\hq}{\hat{q}}
\newcommand {\hx}{\hat{x}}
\newcommand {\hSE}{\hat{\Sigma}_E}
\newcommand {\hPhi}{\hat{\Phi}^t}

\newcommand {\hl}{\hat{\lambda}}
\newcommand {\PP}{\Pi}         
\newcommand {\cc}{\pi}         
\newcommand {\Pix}{{\bf{\Pi}}} 
\newcommand {\bv}{{\bar{v}}}

\newcommand{\rstr}{{\!\hbox{
$\vert\mkern-4.8mu\hbox{\rm\`{}}\mkern-3mu$}}}
\begin{document}
\title {Motion in Periodic Potentials}
\author{Joachim Asch\thanks{CPT-CNRS, Luminy Case 907, 
F-13288 Marseille Cedex 9, France. e-mail: asch@cpt.univ-mrs.fr}
\and 
Andreas Knauf\thanks{Max-Planck-Institute for Mathematics in the Sciences, 
Inselstr.\ 22--26, D-04103 Leipzig, Germany. 
e-mail: Andreas.Knauf@mis.mpg.de}
}
\date{May 1997}
\maketitle
\begin{abstract}
We consider motion in a periodic potential in a classical, quantum,
and semiclassical context. Various results on the distribution
of asymptotic velocities are proven.
\end{abstract}
\section{Introduction}
For a function $V:\bR^d\to\bR$ which is periodic on a regular lattice 
$\cL\subset\bR^d$
we study the evolutions
\beqn
\label{Schr}
i\hbar\partial_t W(t) &=& H^\hbar W(t),\ W(0)=Id \quad \hbox{on } L^2(\bR^d),\\
\label{Ham}
\partial_t\Phi^t &=& X_H\circ\Phi^t,\ \Phi^0=Id\quad \hbox{on }\bR^{2d},
\eeqn
where
$H^\hbar = -{\hbar^2\over2}\Delta+V(\q)$, $H(\p,\q)={\p^2\over2}+V(\q)$ 
and $X_H(\p,\q)=(-\nabla V(\q),\p)$.

\noindent This is done in the limits $t\to\infty$ and $\hbar\to0$.

Why should one get interested in well-known things?
In view of currently very active research on transport (anomalous or not)
in condensed matter physics it is firstly desirable to put on a firm mathematical ground the folklore that `motion in crystals is ballistic';
secondly one should try to obtain at least semiclassical information on quantities 
like the {\em distribution of asymptotic velocities}.

We investigate the asymptotic
velocity $\lim_{t\to\infty}{q(t)/ t}$ and
the asymptotic behaviour of ${q^2(t)/ t^\alpha}$ where  $\alpha=2$
characterizes by definition ballistic, $\alpha=1$ diffusive motion.

Our results and the skeleton of the article are as follows:\\[2mm]
In Sect.\ {\bf 2} we show for a large class of $V$ that the quantum motion is 
ballistic (Theorem \ref{quantumballistic}). That
class of potentials is not optimal in view of singularities. It includes, however,
the Coulomb case. The modulus of the asymptotic velocity is bounded from
above in a natural way (Corollary \ref{maximalvelocity}).

In Sect.\ {\bf 3} we treat the classical motion in smooth potentials.
In $d=2$ dimensions the motion is ballistic
for high enough energies $E$; 
for $d\ge3$ this is true for initial conditions
outside a set of measure $\sim1/\sqrt{E}$.
There always exist fast orbits (of speed $\sim\sqrt{E}$), 
with a dense set of directions.
This is even true for the ergodic case, where the asymptotic speed
is zero with probability one, whereas almost all orbits are unbounded.
(Theorem \ref{thm:cl1}).

The motion is never Anosov (Theorem \ref{nonanosov}).

In particular this means that a gas of particles in a `periodic' container is 
not Anosov if the 
interactions are smooth. So there may be small regions of regular  motions and it seems unlikely that such a gas is ergodic.

To the contrary the planar motion in
periodic potentials with Coulombic ($-1/r$ type) singularities
is known to be of Anosov type and diffusive, \cite{Kn1}. 
In Sect.\ {\bf 4} 
we show that the distribution of asymptotic velocities
--which are  zero on a set of full measure-- is dense 
in a disk of approximate radius $\sqrt{E}$.
(Theorem \ref{densevelocities}).

Concerning semiclassics we show in Sect.\ {\bf 5} that the quantum asymptotic 
velocities are always 
contained in a thickened convex hull of the classical ones for $\hbar$ 
small;
they concentrate in measure inside the convex hull of the support of
the classical probability distribution (Thm.\ \ref{semi:two}).

In particular in the classically ergodic case 
the positive speed of the quantal
motion is only a quantum fluctuation vanishing in the semiclassical limit. 
The same is known also to be true for Coulombic potentials, see \cite{Kn2}. 

The above results are basically consequences of the Birkhoff Ergodic Theorem.
This is interesting insofar our technique is likely to be applicable 
in  different  semiclassical situations.
On the other hand more specific information is needed to prove sharper
results concerning the distribution of asymptotic velocities.

This is done in Sect.\ {\bf 6}, where we consider
separable potentials. There we have fast semiclassical
convergence to the classical velocity distribution
(Theorem \ref{d1semiclassics}).\\[5mm]
{\bf Acknowledgements.} A.K.\ thanks the CPT in Marseille and the members
of the IHES (Bures) for their hospitality during his
visit in May 1997.
\section{Quantum Ballistic Motion}
Now we shall prove that the evolution of a quantum  system in a rather general 
periodic medium is ballistic and  that the asymptotic  velocity exists.
The latter is related to the band functions.

It is known that for a certain class of singular potentials the spectrum 
of the Hamiltonian is absolutely continuous, see Thomas \cite{thom}, Reed-Simon
\cite{RSiv}, Knauf \cite{Kn2}. It is  a folk conjecture that absolute continuity
implies ballistic motion. Our proof in $d$ dimensions is based on Bloch theory.

We consider potentials $V:\bR^d\ar\bR$ which are periodic w.r.t.\ a regular
lattice $\cL\subset\bR^d$ 
\[ V(\q+\ell) = V(q)\qquad (\q\in\bR^d, \ell\in\cL), \]
so that we may consider it as a function $V:\bT\ar\bR$ on the 
unit cell $\bT := \bR^d/ \cL$, and calculate its Fourier transform 
$\cF V:\cL^*\ar\bC$.
Our assumptions on the regularity of the potential are:

\begin{enumerate}
\item[{\bf ($A_q$)}:]
\begin{description}
\item[$d=2$: ] $V\in L^p(\bT)$ with $p>1$, 
\item[$d=3$: ] $V\in L^2(\bT)$ and
\item[$d>3$: ] ${\cF} (V)\in l^p(\cL^*)$ with $p<(d-1)/(d-2)$.
\end{description}
\end{enumerate} 

{\bf ($A_q$)} implies that $V$ is form small with respect to $-\Delta$ and even
operator small for $d\neq 2$, see \cite{CFKS}.

\[\hbox{Denote}\quad D:=-i\hbar\nabla,
\hbox{ then}\qquad H^\hbar:={D^2\over2}\dot+V\] 
is defined by its quadratic form
with form domain
$Q(H^\hbar)=Q(-\Delta)=H^1(\bR^d)$; for $d\neq2$ the operator domain is
$D(H^\hbar)=H^2(\bR^d)$. 

We denote by 
\[O(t):=W^*(t)OW(t)\] 
(with the solution $W(t):=\exp(-iH^\hbar t/\hbar)$ of (\ref{Schr})) the 
Heisenberg time evolution of an operator $O$.
 
The symmetries of $H^\hbar$ allow for a decomposition with respect to the group of 
lattice translations: Let $\cL^\ast$ be the dual lattice with unit cell 
$\bT^{*}$ and denote by
$$U:L^2(\bR^d)\to L^2 \l( \bT^{*},{dk\over\vert\bT^{*}\vert};
      L^2(\bT,{dq})\r)\equiv 
      \int^\oplus_{\bT^*}L^2(\bT,{dq}) {dk\over\vert\bT^{*}\vert}$$
the unitary operator defined by extension from Schwarz space of
$$U\psi(k,q) \equiv (U\psi)_k(q) 
:=\sum_{\ell\in\cL}e^{-ik(q+\ell)}\psi(q+\ell)\qquad
    (\psi\in{\cal S}(\bR^d)).$$
The following facts are known in the literature and will be used below:
\begin{theorem}\label{properties}
Let $V$ satisfy $\bf (A_q)$. Then
\begin{enumerate}
\item
  $UH^\hbar U^{-1}=\int_{\bT^*}^\oplus H^\hbar(k){dk\over\vert\bT^{*}\vert}$ with
  $H^\hbar(k)={1\over2}(D+ \hbar k)^2\dot+ V$ on $L^2(\bT)$,\\ 
  with form domain $Q(H^\hbar(k)) = H^1(\bT)$;
\item
 $k \mapsto H^\hbar(k)$ is a Type (B) analytic family;
\item
  the spectrum of $H^\hbar$ is absolutely continuous;
\item
 $H^\hbar(k)$ has compact resolvent, $H^\hbar(k)=\sum_{n=1}^\infty
E_n^\hbar(k)P_n^\hbar(k)$ where
$E_n^\hbar(k)$ are the eigenvalues in ascending order, $P_n^\hbar(k)$ the
eigenprojections;
\item
 for every $n$ the following are Lebesgue Nullsets:
\[\{k\in\bT^{*}\mid E_n^\hbar\hbox{ is not differentiable at }k\},\]
\[\{k\in\bT^{*}\mid  P_n^\hbar\hbox{ is not differentiable at }k\},\]
\[\{k\in\bT^{*}\mid  \nabla_k E_n^\hbar(k)=0\}.\]
\end{enumerate}

\end{theorem}

{\bf Proof.} 
(1-4)
  are proven in \cite{RSiv}, resp. in \cite{Kn2} for $d=2$.
(5)
  is proven in \cite{Wilc}, \cite{thom}, see also
\cite{GeraNier}.
\hfill $\Box$
\begin{remark} 
We emphasize that while the assumptions $\bf (A_q)$ are sufficient for absolute 
continuity, for $d>2$ they are far from necessary for self-adjointness.
It would be interesting to understand what happens in the gap!
\end{remark}

The result on ballistic transport in the quantum case is
(see also the recent article \cite{GeraNier} by Gerard and Nier):

\begin{theorem}\label{quantumballistic}
Let $V$ satisfy $\bf (A_q)$. It holds
  for $\psi$ with 

$(D\psi,D\psi)+(q\psi, q\psi)<\infty$:
\beqn
\bar{D}\psi:=
\lim_{t\to\infty}{q(t)\psi\over t}=
U^{-1}\l( \int_{\bT^*}^\oplus \sum_{n=1}^\infty P_n(k)(D+ \hbar k)P_n(k)
         \ {dk\over\vert\bT^{*}\vert} \r) U\psi \nonumber \\
=U^{-1}\l( \int_{\bT^*}^\oplus \sum_{n=1}^\infty \hbar^{-1}\nabla_k E_n P_n(k)
         \ {dk\over\vert\bT^{*}\vert} \r) U\psi\nonumber;
\eeqn
  and:
\[\lim_{t\to\infty}  \frac{(\psi, q^2(t)\psi)}{t^2} =
\int_{\bT^{*}}^\oplus \sum_{n=1}^\infty \vert \hbar^{-1}\nabla_k E_n\vert^2
    \Vert P_nU\psi(k)\Vert^2_{L^2(\bT)}\ {dk\over\vert\bT^{*}\vert}>0.\]

\end{theorem}
{\bf Proof.} 
The map 
$$\psi\mapsto{1\over T}\int_0^T D(t)\psi\ dt
   -U^{-1}\l(\int_{\bT^*}^\oplus \sum_{n=1}^\infty P_n(k)(D+ \hbar k)P_n(k)
         \ {dk\over\vert\bT^{*}\vert}\r)U\psi$$
is uniformly bounded from $H^1\to L^2$, consequently it is sufficient
to prove the assertion for $\psi$ in a dense set; indeed firstly  by form smallness of 
$V$ we have the estimate
$$\Vert \langle D\rangle\psi\Vert^2=\langle\psi,(1+D^2)\psi\rangle
\le c_1\vert \langle\psi,H\psi\rangle\vert+c_2\Vert\psi\Vert^2
\le c_3\langle\psi,(1+D^2)\psi\rangle$$
so 
$$\Vert {1\over T}\int_0^T D(t)\psi\ dt\Vert
\le\Vert\langle D\rangle (H+i)^{-1/2}\Vert\Vert(H+i)^{1/2}\psi\Vert
\le c\Vert\psi\Vert_{H^1},$$
secondly
$$
\l\Vert U^{-1}(\int_{\bT^*}^\oplus \sum_{n=1}^\infty P_n(k)(D+ \hbar k)P_n(k)
         \ {dk\over\vert\bT^{*}\vert})U\psi\r\Vert^2\hfill $$
$$=\int_{\bT^{*}}\sum_{n=1}^\infty\l\Vert P_n(k)(D+ \hbar k)(H(k)+i)^{-1/2}P_n(k)
U(H+i)^{1/2}\psi(k)\r\Vert_{L^2(\bT)}^2\ {dk\over\vert\bT^{*}\vert}\hfill $$
$$ \le\int_{\bT^{*}}\l\Vert( D+\hbar k)(H(k)+i)^{-1/2}\r\Vert^2
                \l\Vert U(H+i)^{1/2}\psi(k)\r\Vert^2
    \ {dk\over\vert\bT^{*}\vert}\le const\l\Vert\psi\r\Vert_{H^1}^2.$$
Let $\psi$ such that  
\[U\psi= \l(\int_{\bT^*}^\oplus \sum_{n=1}^N P_n(k)
         \ {dk\over\vert\bT^{*}\vert}\r) U\psi.\]
The set of these is dense in $H^1(\bR^d)$.
Then
\beqno
\lefteqn{\l\Vert U\l({1\over T}\int_0^T D(t)\psi\ dt\r)
   -\l(\int_{\bT^*}^\oplus \sum_{n=1}^N P_n(k)(D+ \hbar k)P_n(k)
         \ {dk\over\vert\bT^{*}\vert}\r) U\psi\r\Vert^2}\\
&\hspace{-2cm}\le& \hspace{-1.1cm}\int_{\bT^{*}}\l\Vert \sum_{m,n}^{\infty,N}
   {1\over T}\int_0^T \exp\l({i(E_m(k)-E_n(k))t/\hbar}\r) dt P_m(k)(D+ \hbar
k)P_n(k)U\psi(k)\r.\\
&\hspace{-2cm}& 
\l. - \sum_{n=1}^N P_n(k)(D+ \hbar k)P_n(k)U\psi(k)\r\Vert^2_{L^2(\bT)}
          \ {dk\over\vert\bT^{*}\vert}\\
&\hspace{-2cm}=&
\hspace{-1cm}\int_{\bT^{*}}\sum_{\stackrel{m=1}{m\neq n}}^\infty \l\Vert P_m(k)
\sum_{n=1}^N{1\over T}\int_0^T e^{i(E_m(k)-E_n(k))t/\hbar}\ dt 
       (D+ \hbar k)P_n(k)U\psi(k)\r\Vert^2\ \!\!\!\!\!
       {dk\over\vert\bT^{*}\vert}\\
&\hspace{-2cm}&\hspace{-1cm} \to0\qquad(T\to\infty) 
\eeqno 
by dominated convergence, which is applicable because \\
$\Vert P_m(k)\sum_n\ldots\Vert=O(1/T)$ for $m\neq n$, almost all $k$,
and is uniformly majorized by
$$\mbox{const}\cdot 
\sup_{n=1,\ldots,N}\Vert P_m(k)(D+ \hbar k)P_n(k)U\psi(k)\Vert^2_{L^2(\bT)}$$
which is summable with respect to $m$ and $k$. 

In \cite{RadiSimo} it was shown that for
$\psi\in H^1(\bR^d)\cap D(\vert q\vert)$:
$$q(T)\psi = q\psi+
\int_0^T D(t)\psi dt. $$
It follows
\beqno
\lim_{T\to\infty}{q(T)\psi\over T}
&=& \lim_{T\to\infty} {1\over T}\int_0^T D(t) \psi dt  \\
&=& U^{-1}\l(\int_{\bT^*}^\oplus \sum_{n=1}^\infty P_n(k)(D+ \hbar k)P_n(k)
         \ {dk\over\vert\bT^{*}\vert} \r) U\psi.
\eeqno
It remains to establish the identity
$$P_n(k)(D+ \hbar k)P_n(k)= \hbar^{-1}\nabla_k E_n(k)P_n(k)$$
for almost every $k\in\bT^*$. But this follows from:\\ 
$\nabla_k PHP=E(\nabla_k P)P+EP(\nabla_k P)+P(\nabla_k H)P=
E\nabla_k P+P(\nabla_k H)P$, \\
$\nabla_k PHP=E\nabla_k P +P\nabla_k E$, $\nabla_k H=\hbar(D+ \hbar k)$, all valid
in the quadratic form sense for almost every $k$ by Properties \ref{properties}.

This was the first assertion; the second one is a
consequence thereof. The positivity is inferred from Theorem 
\ref{properties}.
\hfill $\Box$

\medskip
As a corollary we get an estimate for the group velocity in 
one band:

\begin{coro} \label{maximalvelocity}
For every $n$ there is a set of full measure of $k$'s such that
$$\l\vert {\nabla_k E^\hbar_n(k)\over\hbar}\r\vert^2\le
2(E^\hbar_n(k)-\inf_{\{\Vert\psi\Vert=1,P_n(k)\psi=\psi\}}
\langle\psi,V\psi\rangle_{L^2(\bT)})$$
\end{coro}
\noindent{\bf Proof.} Let $\psi\in L^2(\bT)$, $k$ such that $E_n,P_n$ are
differentiable at $k$.
\beqno
\hspace{-5mm}\vert \hbar^{-2} \nabla_k E_n(k)\vert^2\Vert P_n(k)\psi\Vert^2&=&
\Vert P_n(k)(D+ \hbar k)P_n(k)\psi\Vert^2\\
&\le& \Vert (D+ \hbar k)P_n(k)\psi\Vert^2\\
&=& 
2(E_n(k)\Vert P_n(k)\psi\Vert^2-\langle P_n(k)\psi,V P_n(k)\psi\rangle)
\eeqno
implies the inequality.
\hfill $\Box$
\section{Classical Motion: Smooth Potentials} \label{sect:cl:sm}
The classical motion in a $\cL$--periodic potential $V$ on $\bR^d$
is described by Hamilton's equations (\ref{Ham}) on phase space
$P := T^*\bR^d$ for $H:P\ar \bR$, 
$H(p,q) = \eh p^2 +V(q)$. If $V\in C^2(\bR^d,\bR)$ 
(as we assume in this section), 
the flow $\Phi^t : P \ar P$
exists uniquely for all times $t\in\bR$.

We will analyze its restrictions
$\Phi^t_E := \Phi^t\rstr_\SE$ to the energy shells 
\[\SE := H^{-1}(E).\] 

Alternatively we study motion on the phase space $\hP:=T^*\bT$
over the configuration torus (and mark corresponding objects with a hat).
Using the phase space projection $\PP:P\ar\hP$ arising
from the projection $\cc:\bR^d\ar\bT=\bR^d/\cL$ of configuration spaces,
we thus consider the flow $\hPhi:\hP\ar\hP$ 
generated by the Hamiltonian function
$\hH:\hP\ar\bR$, $\hH\circ \PP=H$, and 
its compact energy shells $\hSE:=\hH^{-1}(E)$ with the restricted flows
$\hPhi_E := \hPhi\rstr_{\hSE}$.

The Liouville measures $\hl$ of the phase space regions $\hH^{-1}([\Vmin,E])$
$E\in\bR$, are now finite, a fact which enables us to use notions of ergodic 
theory.

The energy scales 
$\Vmin := {\rm inf}_{q\in \bR^d} V(q)$,
\[\Vmean := \int_{\bT} V(q)dq/|\bT|\] 
and $\Vmax := {\rm sup}_{q\in \bR^d} V(q)$ of the dynamics will be used
repeatedly.

As a consequence of Birkhoff's Ergodic Theorem for $\hl$--almost all 
$\hx_0\in\hP$
\[\bv^\pm(\hx_0) := \lim_{T\ar\pm\infty} \frac{1}{T} \int_0^T \hp(t,\hx_0) dt \]
exist and are equal. In this case we set $\bv:=\bv^\pm$, and otherwise
$\bv:=0$, thus defining the {\em asymptotic velocity}
\[\bv: \hP\ar\bR^d\]
which is a measurable phase space function. 

We denote its lift $\bv\circ\PP:P\ar\bR^d$ to the original phase space $P$ by
the same symbol and thus have
\[\lim_{t\ar\pm\infty} \frac{\q(t,x_0)}{t} = \bv(x_0)\] 
$\lambda$--almost everywhere. 

$\hPhi$ is called {\em ballistic at} $\hx\in\hP$ if $\bv(\hx)\neq 0$
(observe that by the above definition this implies existence and equality
of $\bv^\pm$). 

We are particularly interested in the energy dependence of
asymptotic velocity and thus introduce
the  {\em energy-velocity map} 
\beq
A := (\hH,\bar{v}): \hP \ar \bR^{d+1}.
\label{def:A}
\eeq	
$A$ is measurable and generates an image measure
$\nu := \hl A^{-1}$ on $\bR^{d+1}$.\\[2mm] 
{\bf Example.} 
In the simplest case $V=0$ of {\em free motion} $\nu$
is a smooth measure on the paraboloid
\[A(\hP) = \{(\eh v^2,v)\mid v\in\bR^d\}.\]
\medskip
Like in the above example, in the general case
$\nu$ is invariant under $(h,v)\mapsto(h,-v)$, since
the motion is reversible. 

The equality $H=\eh\bv^2$ is in the general case replaced by the estimate
$|\bv(x)| \leq \sqrt{2(H(x)-V_{\min})}$.

For regular values $E$ of the energy 
one may consider the probability 
distribution of the asymptotic velocities $\bv$ 
w.r.t.\ the normalized Liouville measure
$\hl_E$ on the energy shell $\hSE$. By the above bound this is supported
within a ball of radius $\sqrt{E-\Vmin}$.  

Unlike in the above example in general it
is not expected to depend weak--*--continuously on $E$,
See Remark \ref{rema}.\ref{one} below.

Here are our results on classical ballistic motion for $V\in C^2$--potentials:
\begin{theorem} \label{thm:cl1}
\begin{enumerate}
\item
For $d=1$ the motion is ballistic at $x=(p,q)\in P$ iff 
$E:= H(x) > \Vmax$, with 
asymptotic velocity
\[ \bar{v}(x) = 
\frac{{\rm sign}(p)}{l^{-1} \int_0^l (2(E-V(q)))^{-\eh}dq}\] 
($l>0$ being the period of $\cL$).
\item \label{th:dense}
For $d>1$ and $E > \Vmax$ there exists a set $B_E\subset\SE$ 
for which the motion is ballistic, whose directions
\[ \{\bar{v}(x)/\|\bar{v}(x)\|\mid x\in B_E\} \]
are dense in $S^{d-1}$, with moduli
\beq 
\frac{\sqrt{2}(E-\Vmax)} {\sqrt{E-\Vmean}} 
\leq \|\bar{v}(x)\| \leq \sqrt{2(E-\Vmin)},\qquad (x\in B_E).
\label{moduli}
\eeq
\item
For $d=2$ and $V\in C^5(\bR^d,\bR)$ there exists a threshold $\Eth\geq\Vmax$
above which the flows
$\Phi^t_E$ ($E>\Eth$) are ballistic $\hl_E$--almost everywhere.

$\Eth$ is given by the following condition.
For $E>\Eth$ there are two geometrically different minimal
tori $\bT^2_1,\bT^2_2\subset\hSE$ (by `geometrically different' we mean:
not related by time reversal symmetry $\cI(\hp,\hq) := (-\hp,\hq)$).
\item
We assume here that $V$ is $3d$ times continuously differentiable.
Then for $d > 2$ there exist a threshold energy $\Eth\geq\Vmax$ 
and for $E>\Eth$ subsets $\hat{B}_E \subset\hSE$ of measures
\[\hl_E(\hat{B}_E)\geq 1-\sqrt{\Eth/E}\]
such that on $\hat{B}_E$ the motion is ballistic.
\item \label{th:erg}
If the flow $\hPhi_E$ on the energy shell is ergodic w.r.t.\ $\hl_E$, 
then $\bv=0$ with $\hl_E$--probability one. However, if in addition $E>\Vmax$, 
the trajectories are unbounded with probability one:
\[\hl_E \l( \l\{ \hat{x}_0\in\hSE\l|\, 
\limsup_T\l\|\int_0^T\hat{p}(t,\hat{x}_0)dt\r\| = \infty \r. \r\}\r)=1.\] 
\item
For $d\geq 2$ there are smooth $\cL$--periodic potentials $V$ and energies 
$E>\Vmax$ whose energy shell contains a set of bounded orbits of positive measure:  
\[\hl_E \l( \l\{ \hat{x}_0\in\hSE\l|\, 
\limsup_T\l\|\int_0^T\hat{p}(t,\hat{x}_0)dt\r\| < \infty \r. \r\}\r) >0.\] 

\end{enumerate}
\end{theorem}
{\bf Example.} 
Consider first in $d=1$ dimensions the potential $V(q)=\cos(q)$.
With the formula of Thm.\ \ref{thm:cl1} for $E\ge 1$ the asymptotic speed
equals
\[\bar{v}(E) = \frac{\pi\sqrt{E-1}}{\sqrt{2}Elliptic{\cal K}(2/(1-E)) } ,\]
($Elliptic{\cal K}$ being the complete elliptic integral of the first kind) 
and $\bar{v}(E)=0$ for $-1\le E\le 1$.

\begin{figure} 
\centerline{}
\epsfig{file=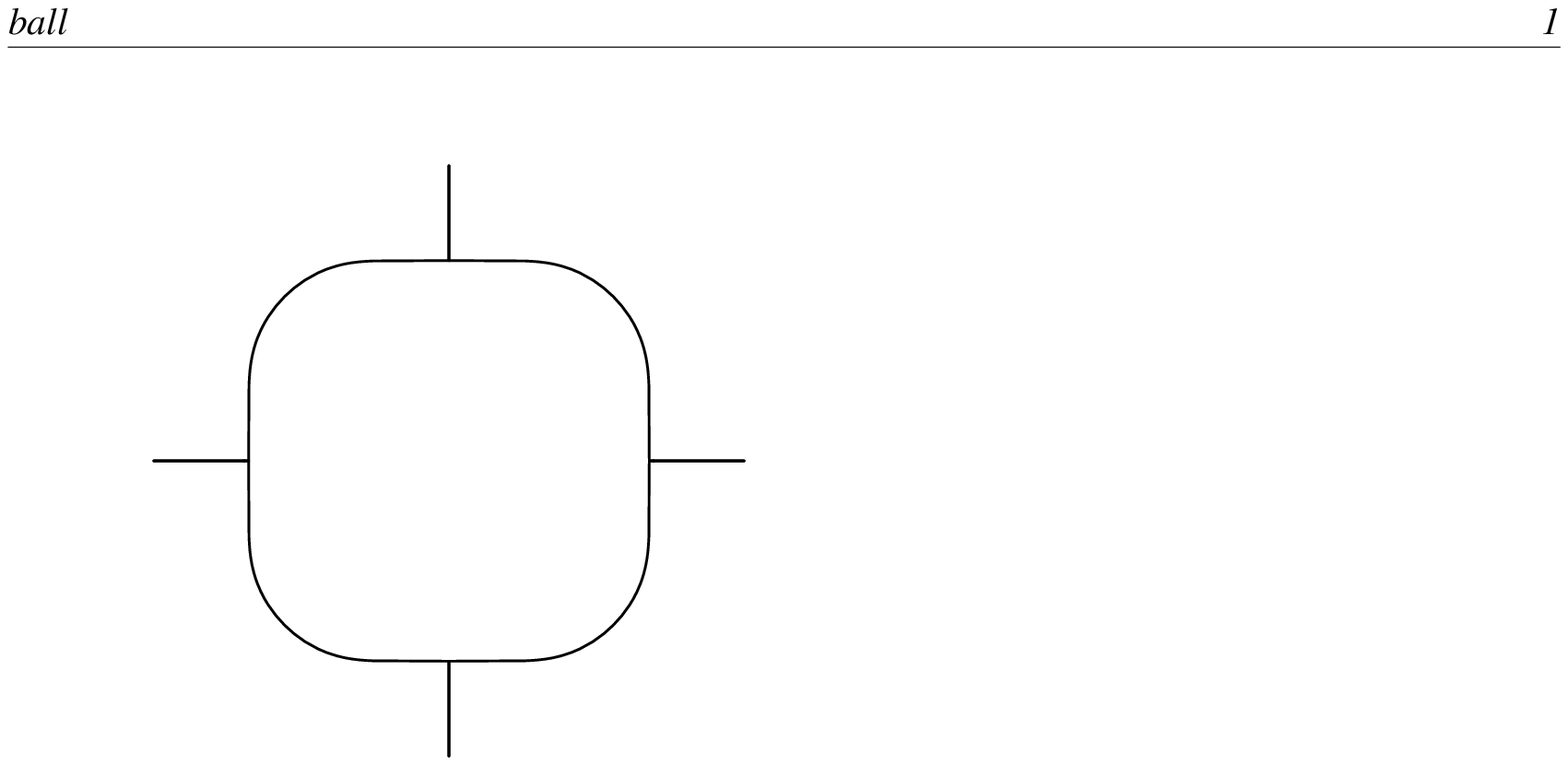,height=5cm,bbllx=90,bblly=510,bburx=300,bbury=710, clip=}
\caption{Distribution of asymptotic velocities $\bar{v}$ for the separable potential 
$V(q)=\cos(q_1)+\cos(q_2)$ and energy $E=3$}
\label{figure1}
\end{figure}

For $d=2$ this leads to a distribution of asymptotic velocities
for energy $E$
of the potential $V(q)=\cos(q_1)+\cos(q_2)$ depicted in Figure \ref{figure1}.
Observe that there is a positive probability for motion along the axes.\\[2mm]
{\bf Proof.} 
\begin{enumerate}
\item 
$\bv(x_0) = l/T$ where $T := \int_0^l \frac{dq}{\dot{q}}$ is the time
needed for the spatial period $l$.
\item
The idea is to construct periodic but non-contractible orbits on the torus.
These are covered by ballistic orbits in the configuration space $\bR^d_q$.
  
For $E > \Vmax$ we consider the geodesic motion on $\bT$ in the 
{\em Jacobi metric} 
\beq
\gEh(\hq) := (E-V(\hq))\cdot \sum_{i=1}^d d\hq_i\otimes d\hq_i.
\label{Jacobi}
\eeq
The geodesics of that metric are known to coincide with the solution curves
$t\mapsto \hq(t,\hx_0)$, $\hx_0\in \hSE$, up to a time reparametrization 
$\tau\mapsto t(\tau)$.  

In every nontrivial homotopy class $l$ of the fundamental group 
$\pi_1(\bT)\cong\cL$ 
we find a shortest closed geodesic $\hat{c}:S^1\ar\bT$ (with $S^1:=\bR/\bZ$). 
The length
\beq
L(\hat{c})=\int_0^1 \l\|\frac{d\hat{c}(\tau)}{d\tau}\r\| \cdot\sqrt{E-V(\hat{c}(\tau))}d\tau
\label{J:length}
\eeq
($\l\|\cdot\r\|$ denoting Euclidean norm)
of that geodesic in the Jacobi metric is the infimum of the lengths of the curves in its homotopy class $l$. 
Let the corresponding solution curve $t\mapsto \hat{q}(t,\hat{x}_0)$
on the torus have period $T$. 

We have $q(nT,x_0) = q(0,x_0)+n\cdot l$ for motion in configuration space 
$\bR^d_q$, starting from a point $x_0\in \PP^{-1}(\hx_0)$.
Therefore the asymptotic velocity $\bar{v}(x_0)$ of this orbit
exists and equals $\bar{v}(x_0) = l/T$. 
So our task is to estimate $T$ from below and above.

The upper bound
\[\|\bar{v}(x)\|\leq \sqrt{2(E-\Vmin)}\]
follows from the general bound $\|p\| \leq \sqrt{2(E-\Vmin)}$ if 
$(p,q)\in\SE$.

In order to prove the lower bound for $\|\bv\|$, 
we derive an upper bound for the period $T$ and argue as follows. The length
$L(\hat{c})$ of our minimal geodesic $\hat{c}$ 
is shorter than the lengths of all the homotopic straight lines 
$\tilde{c} \equiv \tilde{c}_{\hq}: S^1\ar\bT$, 
\[\tilde{c}(\tau) := \hq+\tau\cdot l\ ({\rm mod}\ \cL)\] 
starting from a point $\hq\in\bT$.
But by formula (\ref{J:length}) the length 
$L(\tilde{c}) = \|l\| \cdot \int_0^1 \sqrt{E-V(\tilde{c}(\tau))}d\tau$.
By concavity of $x\mapsto \sqrt{E-x}$ 
\[\int_0^1 \sqrt{E-V(\tilde{c}(\tau))}d\tau \leq \sqrt{E-\Vmean}\]
for some $\hq$. So we obtain
\beq
L(\hat{c})\leq L(\tilde{c}) \leq \|l\| \cdot \sqrt{E-\Vmean}.
\label{LL}
\eeq
On the other hand the period 
\beqno
T &=& \int_0^1 \frac{dt}{d\tau} d\tau = 
\int_0^1  \l\|\frac{d\hat{c}(\tau)}{d\tau}\r\| /\sqrt{2(E-V(\tilde{c}(\tau)))}d\tau\\
&\leq& L(\hat{c})/(\sqrt{2}(E-\Vmax)). 
\eeqno
Together with (\ref{LL}) this gives 
$T\leq \|l\| \cdot \sqrt{E-\Vmean}/(\sqrt{2}(E-\Vmax))$
from which the lower estimate for the asymptotic speed 
$\|\bar{v}(x_0)\| = \|l\| /T$ in (\ref{moduli}) follows.

The asymptotic {\em direction} $\bv(x_0)/\|\bv(x_0)\|$ of our ballistic orbit
equals $l/\|l\|$.   
But the directions of lattice points $l\in\cL\setminus\{0\}$, seen from the origin are dense in $S^{d-1}$.
\item
We show that under the existence assumption for the tori 
$\bT_1,\bT_2\subset\hSE$ the motion is ballistic $\hl_E$--a.e. 

For $E>\Vmax$ and $d=2$ the energy shell $\hSE$ is diffeomorphic to 
$S^1\times \bT^2$, $S^1$ representing the circle of directions $\hp/\|\hp\|$.

By the minimality condition the tori 
\[\bT_1,\bT_2\subset\hSE\subset\hP\sim\bR^2\times\bT\] 
(and their time inverses)
$\cI(\bT_1)$, $\cI(\bT_2)$) project diffeomorphically to the configuration torus
$\bT$. Thus we may represent them as graphs of functions 
$\hat{P}_1,\hat{P}_2:\bT\ar\bR^2$. For $i=1,2$ the mapping 
\[  \hat{P}_i/\|\hat{P}_i\|:\bT\ar S^1 \]
are the local direction, and is by minimality topologically trivial.
So their complement 
\[\hSE-(\bT_1\cup\bT_2\cup\cI(\bT_1)\cup\cI(\bT_2))\]
in the energy shell 
of four components diffeomorphic to thickened two-tori. 
These components 
roughly correspond to sectors of directions in which the particle is forced to move.

The problem which has to be overcome is that these sectors of directions depend
on the point $q$, and that their union for all $q$ may have a total opening 
angle of more than $\pi$. Thus it may happen that the particle
goes backward for some time. 

Without loss of generality we consider {\em the} component 
$\hat{C}\subset\hSE-(\bT^2_1\cup\bT^2_2)$ which consists of points 
$(\hp,\hq)\in\hSE$
which are linear combinations 
$\hp = \alpha_1 \hat{P}_1(\hq) + \alpha_2 \hat{P}_2(\hq)$ 
of the points $(\hat{P}_i(\hq),\hq)\in\bT^2_i$ 
with {\em positive} coefficients $\alpha_i$.

On the invariant tori $\bT^2_i$ the motion 
is conditionally periodic with frequency vectors
$\om_i\in\bR^2\setminus\{0\}$. 
If we consider the Lagrangian manifolds $M_i\subset\SE$ which under $\PP$
project to the tori $\bT^2_i$, 
these manifolds are not only diffeomorphic to $\bR^2$,
but by our minimality assumption they project under $\Pix: P \ar \bR^2$, 
$(p,q)\mapsto q$ 
diffeomorphically to the configuration plane $\bR^2$. 
Thus they induce two flows
\[\Psi_i^{t}:\bR^2\ar \bR^2, \qquad \Psi_i^{t} := 
\Pix\circ\Phi^t\circ \l(\Pix\rstr_{M_i}\r)^{-1}.\]
These are nearly linear in the sense
\beq
\Psi_i^{t}(q_0)=q_0+\om_i\cdot t + {\cO}(t^0)
\label{nearly:linear}
\eeq
with ${\cO}(t^0)$ uniform in $q_0$, 
since they come from a conditionally periodic motion 
on a torus. The flow lines of $\Psi_1$ and $\Psi_2$ both foliate the configuration plane
and are transversal to each other. Since both foliations project under 
$\cc:\bR^2\ar\bT^2$ 
to foliations of the (compact!) torus, the angles under which these
foliations intersect are bounded away from $0$ and $\pi$, see Figure
\ref{figure2}.
\begin{figure}
\begin{center}
\begin{picture}(0,0)%
\includegraphics{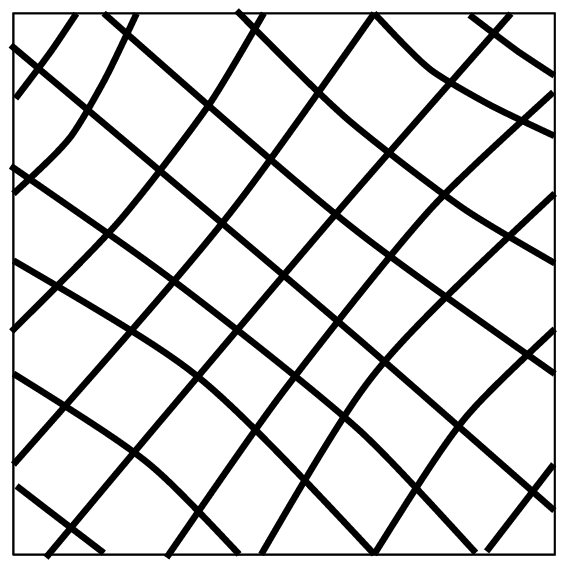}%
\end{picture}%
\setlength{\unitlength}{0.00083300in}%
\begingroup\makeatletter\ifx\SetFigFont\undefined
\def\x#1#2#3#4#5#6#7\relax{\def\x{#1#2#3#4#5#6}}%
\expandafter\x\fmtname xxxxxx\relax \def\y{splain}%
\ifx\x\y   
\gdef\SetFigFont#1#2#3{%
  \ifnum #1<17\tiny\else \ifnum #1<20\small\else
  \ifnum #1<24\normalsize\else \ifnum #1<29\large\else
  \ifnum #1<34\Large\else \ifnum #1<41\LARGE\else
     \huge\fi\fi\fi\fi\fi\fi
  \csname #3\endcsname}%
\else
\gdef\SetFigFont#1#2#3{\begingroup
  \count@#1\relax \ifnum 25<\count@\count@25\fi
  \def\x{\endgroup\@setsize\SetFigFont{#2pt}}%
  \expandafter\x
    \csname \romannumeral\the\count@ pt\expandafter\endcsname
    \csname @\romannumeral\the\count@ pt\endcsname
  \csname #3\endcsname}%
\fi
\fi\endgroup
\begin{picture}(3219,3025)(79,-2485)
\put(2837,-2446){\makebox(0,0)[lb]{\smash{\SetFigFont{12}{14.4}{rm}$q_1$}}}
\put( 79,313){\makebox(0,0)[lb]{\smash{\SetFigFont{12}{14.4}{rm}$q_2$}}}
\end{picture}
\end{center}
\caption{The two foliations of the configuration torus}
\label{figure2}
\end{figure}
However, $\Psi_1$ and $\Psi_2$ do not commute in general. Nevertheless, 
we may use them to find an adapted coordinate system 
\[\Psi:\bR^2\ar \bR^2,\quad 
(s_1,s_2)\mapsto \Psi_1^\bR\circ \Psi_2^{s_2}(q_0)\cap\Psi_1^{s_1}\circ \Psi_2^\bR(q_0),\]   
since the orbit $\Psi_1^\bR\circ \Psi_2^{s_2}(q_0)$ through the point $\Psi_2^{s_2}(q_0)$
has a unique intersection with the orbit $\Psi_1^{s_1}\circ \Psi_2^\bR(q_0)$ through
$\Psi_2^\bR(q_0)$. By the above remarks we have
\[\Psi(s_1,s_2)=q_0+\om_1\cdot s_1+\om_2\cdot s_2 + {\cO}(1), \]
and the Jacobian of $\Psi$ is uniformly bounded.
   
We consider the component $C=\PP^{-1}(\hat{C})$ of 
$\SE$
and an initial point $x_0=(p_0,q_0)\in C$. By compactness 
of $\bT$ the angle between $\hat{P}_1(\hq)$ and $\hat{P}_2(\hq)$
is bounded away from $0$ and $\pi$.
Thus for some $c>0$ each point 
$(p,q)\in C$ has a momentum vector $p$ which is a linear combination 
$p=\alpha_1 p_1+ \alpha_2 p_2$ with $(p_i,q)\in M_i$ and 
$\alpha_1 + \alpha_2\geq c$. 

Thus the $\Psi$--coordinates $s_1,s_2$ are increasing along
the trajectory $t\mapsto q(t,x_0)$, and there exists a
$c'>0$ with $\frac{d}{dt} (s_1(t)+s_2(t))\geq c'$.

The linear coordinate $\tilde{q}:\bR^2\ar \bR$,
$ \tilde{q} := (\om_1/\|\om_1\|+\om_2/\|\om_2\|)\cdot q$
on the configuration plane increases (at least) linearly along the
trajectory. Namely, by (\ref{nearly:linear})
the trajectory meets the inequality 
\beqno 
\lefteqn{\tilde{q}(q(t,x_0))}\\
&=& \l( \frac{\om_1}{\|\om_1\|} + \frac{\om_2}{\|\om_2\|} \r)\cdot
\Psi(s_1(t),s_2(t)\\
&=& \l( \|\om_1\| +\frac{\om_1\cdot\om_2}{\|\om_2\|}\r)s_1(t) +
 \l( \|\om_2\| +\frac{\om_1\cdot\om_2}{\|\om_1\|}\r)s_2(t) + \cO(1)\\
&\geq& c_{II}(s_1(t)+s_2(t)) \geq c_{I}c_{II}\cdot t,
\eeqno
if $t$ is large.
That is, {\em if} the asymptotic velocities $\bar{v}^\pm(x_0)$ exist
and are equal, they must be non--zero.

The existence of such minimal tori for large $E$ follows under our 
differentiability assumption $V\in C^5(\bR^2,\bR)$ 
from the results of the paper \cite{Po} of P\"{o}schel.
\item
In the Jacobi metric (\ref{Jacobi}) the perturbation of the integrable part equals
$(1-V(q)/E)\cdot \sum_{i=1}^d dq_i\otimes dq_i$. Since $\hat{V}\in C^{3d}(\bT,\bR)$,
the norm of the perturbation is finite and proportional to $E^{-1}$.
The unperturbed part of the Hamiltonian function for geodesic motion
in that metric is just the Hamiltonian of free motion. So up to a 
linear transformation the momenta coincide with the action variables
of this integrable system,  and the nondegeneracy condition
of the frequencies is satisfied. These depend analytically on the 
action variables, and the perturbation is $C^{3d}$ and of size
$\vep=\cO(1/E)$
We can apply the result by P\"{o}schel on the measure of KAM
tori for a perturbation of order $\vep$ (Corr.\ 2 of \cite{Po}), 
which says that the 
measure of the complement of the KAM tori is of order $\sqrt{\vep}$. 
\item 
If $\lambda_E(\{\bar{v} = 0\}) < 1$, there exists an index 
$j\in\{1,\ldots,d\}$ with probabilities 
\[\lambda_E(\{ \bar{v}_j > 0 \}) = \lambda_E(\{ \bar{v}_j < 0 \}) > 0.\]
But this contradicts ergodicity, since these two exclusive events are
flow--invariant.

To show that the trajectories are unbounded with 
probability one if $E>\Vmax$, 
we consider the flow-invariant measurable events 
$E_n\subset\hSE$, $n\in\bN$ defined by
\[E_n := \l\{ \hat{x}_0\in\hSE\l|\, 
\exists t_1,t_2\in\bR: 
\l\|\int_{t_1}^{t_2}\hat{p}(t,\hat{x}_0)dt\r\| \geq n \r. \r\}.\] 
We know from part 2 of the theorem that there are ballistic trajectories.
Each point $\hat{x}\in\hSE$ on such a ballistic trajectory
is contained in all the sets $E_n$.
By absolute continuity of $\|\int_{t_1}^{t_2}\hat{p}(t,\hat{x}_0)dt\|$ w.r.t.\ 
the initial condition $\hat{x}_0$ we conclude that for all 
$n\in\bN$ the Liouville measure $\lambda_E(E_n)>0$.
But the flow being ergodic, and $E_n$ flow-invariant,
$\lambda_E(E_n)$ can only zero or one. 
So the set $\cap_{n\in\bN} E_n$ of unbounded trajectories has measure one.  
\item
To construct potentials $V\in C^\infty(\bR^d,\bR)$, $d\geq2$, which have 
energies $E>\Vmax$ with many bounded orbits on the energy shell, one
writes $V(\q):=\sum_{\ell\in\cL} W(\q-\ell)$
with $W\in C^\infty_0(\bR^d,\bR)$, $W(\q):= \tilde{W}(|q|)$, with
$\tilde{W}(r)=0$ for $r\geq \eh\min_{\ell\in\cL} \|\ell\|$, so that the 
supports of the lattice-translated $W$ do not overlap.
So as long as the particle is captured near a lattice point $\ell$,
the motion is one in a potential centrally symmetric around $\ell$, and
the angular momentum around that point is constant.

Thus one reduces the dimension by considering the effective potential 
$\tilde{W}_L(r):=\tilde{W}(r)+\frac{L^2}{2r^2}$ for angular momentum $L$.
For a given choice of $L\neq 0$ one chooses 
$\tilde{W}\leq 0$ so that the effective potential has a strictly 
positive nondegenerate minimum $r_0$:  $\tilde{W}_L(r_0)>0$, 
$\frac{d}{dr}\tilde{W}_L(r_0)=0$, $\frac{d^2}{dr^2}\tilde{W}_L(r_0)>0$.
Then the assertion holds for $E= \tilde{W}_L(r_0) +\vep$, since
$\Vmin\leq 0$.

Whether one can construct for $d\geq3$
potentials with a positive measure of bounded orbits on energy shells of 
{\em arbitrarily large} energy, is a much more complicated question.
\hfill $\Box$
\end{enumerate}
Part 2 and 3 of the above theorem show that the minimal KAM tori 
play an important role in the distribution of asymptotic velocities.
However, it is known \cite{Kn3} that for $d\geq 2$ an energy shell 
can only be foliated by such tori if $V$ is constant.
This suggests that tori which do not diffeomorphically project to
the configuration torus are important for the high energy 
distribution of asymptotic velocity, see Figures \ref{figure3}
and \ref{figure4}.
\begin{figure} 
\begin{center}
\begin{picture}(0,0)%
\includegraphics{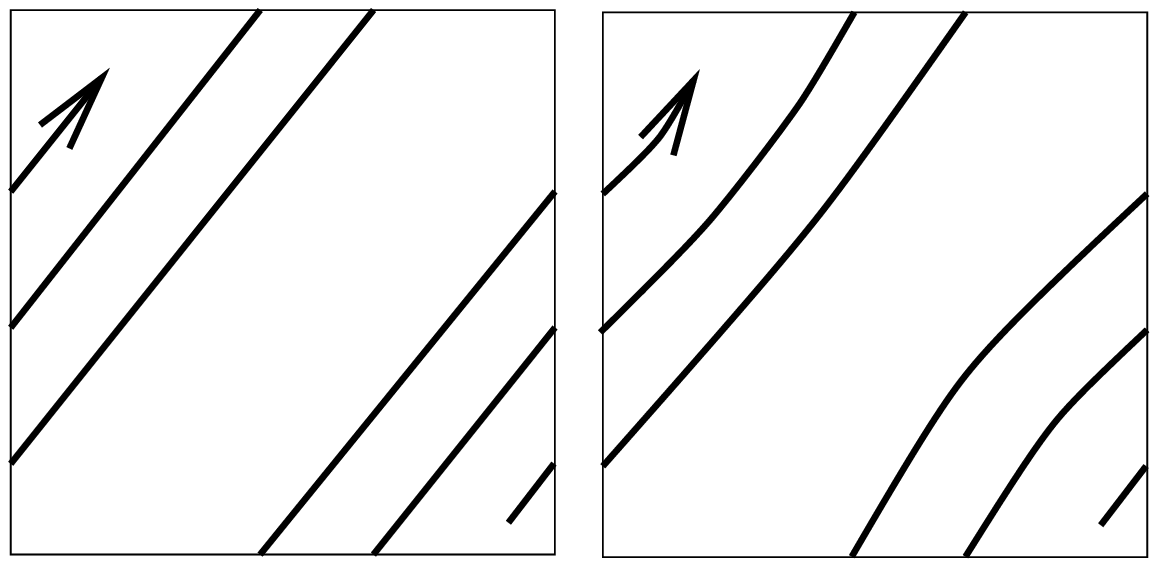}%
\end{picture}%
\setlength{\unitlength}{0.00083300in}%
\begingroup\makeatletter\ifx\SetFigFont\undefined
\def\x#1#2#3#4#5#6#7\relax{\def\x{#1#2#3#4#5#6}}%
\expandafter\x\fmtname xxxxxx\relax \def\y{splain}%
\ifx\x\y   
\gdef\SetFigFont#1#2#3{%
  \ifnum #1<17\tiny\else \ifnum #1<20\small\else
  \ifnum #1<24\normalsize\else \ifnum #1<29\large\else
  \ifnum #1<34\Large\else \ifnum #1<41\LARGE\else
     \huge\fi\fi\fi\fi\fi\fi
  \csname #3\endcsname}%
\else
\gdef\SetFigFont#1#2#3{\begingroup
  \count@#1\relax \ifnum 25<\count@\count@25\fi
  \def\x{\endgroup\@setsize\SetFigFont{#2pt}}%
  \expandafter\x
    \csname \romannumeral\the\count@ pt\expandafter\endcsname
    \csname @\romannumeral\the\count@ pt\endcsname
  \csname #3\endcsname}%
\fi
\fi\endgroup
\begin{picture}(5750,3104)(377,-2501)
\put(377,238){\makebox(0,0)[lb]{\smash{\SetFigFont{12}{14.4}{rm}$q_2$}}}
\put(5777,-2462){\makebox(0,0)[lb]{\smash{\SetFigFont{12}{14.4}{rm}$q_1$}}}
\put(3077,-2462){\makebox(0,0)[lb]{\smash{\SetFigFont{12}{14.4}{rm}$q_1$}}}
\end{picture}

\end{center}
\caption{Motion on unperturbed (left) and perturbed (right) invariant torus}
\label{figure3}
\end{figure}
\begin{figure} 
\begin{center}
\begin{picture}(0,0)%
\includegraphics{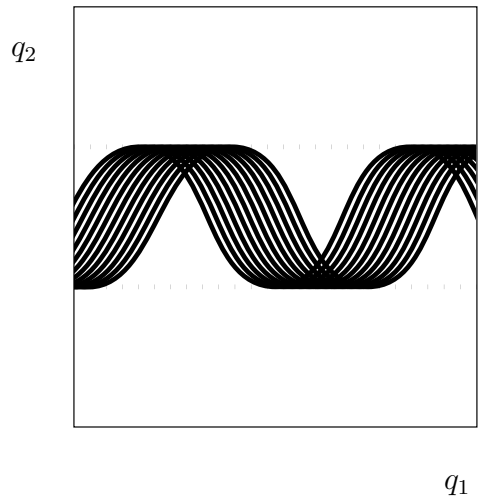}%
\end{picture}%
\setlength{\unitlength}{0.00083300in}%
\begingroup\makeatletter\ifx\SetFigFont\undefined
\def\x#1#2#3#4#5#6#7\relax{\def\x{#1#2#3#4#5#6}}%
\expandafter\x\fmtname xxxxxx\relax \def\y{splain}%
\ifx\x\y   
\gdef\SetFigFont#1#2#3{%
  \ifnum #1<17\tiny\else \ifnum #1<20\small\else
  \ifnum #1<24\normalsize\else \ifnum #1<29\large\else
  \ifnum #1<34\Large\else \ifnum #1<41\LARGE\else
     \huge\fi\fi\fi\fi\fi\fi
  \csname #3\endcsname}%
\else
\gdef\SetFigFont#1#2#3{\begingroup
  \count@#1\relax \ifnum 25<\count@\count@25\fi
  \def\x{\endgroup\@setsize\SetFigFont{#2pt}}%
  \expandafter\x
    \csname \romannumeral\the\count@ pt\expandafter\endcsname
    \csname @\romannumeral\the\count@ pt\endcsname
  \csname #3\endcsname}%
\fi
\fi\endgroup
\begin{picture}(4927,3072)(1496,-2352)
\put(2028,387){\makebox(0,0)[lb]{\smash{\SetFigFont{12}{14.4}{rm}$q_2$}}}
\put(4728,-2313){\makebox(0,0)[lb]{\smash{\SetFigFont{12}{14.4}{rm}$q_1$}}}
\end{picture}
\end{center}
\caption{Motion on an invariant torus which does not project diffeomorphically 
onto the configuration torus}
\label{figure4}
\end{figure}
Indeed, in the separable case $V(q) = \sum V_j(q_j)$ with $V_j$ non-constant, 
the probability to move in each of the directions $\ell_1,\ldots,\ell_d$
is positive (Figure \ref{figure1}).
In the non-separable case like the one depicted in Figure \ref{figure5}, 
it can be proven by perturbation arguments that 
there are lattice-rational directions in which the particle moves with 
positive probability.
\begin{figure}\label{figure5}
\centerline{}
\epsfig{file=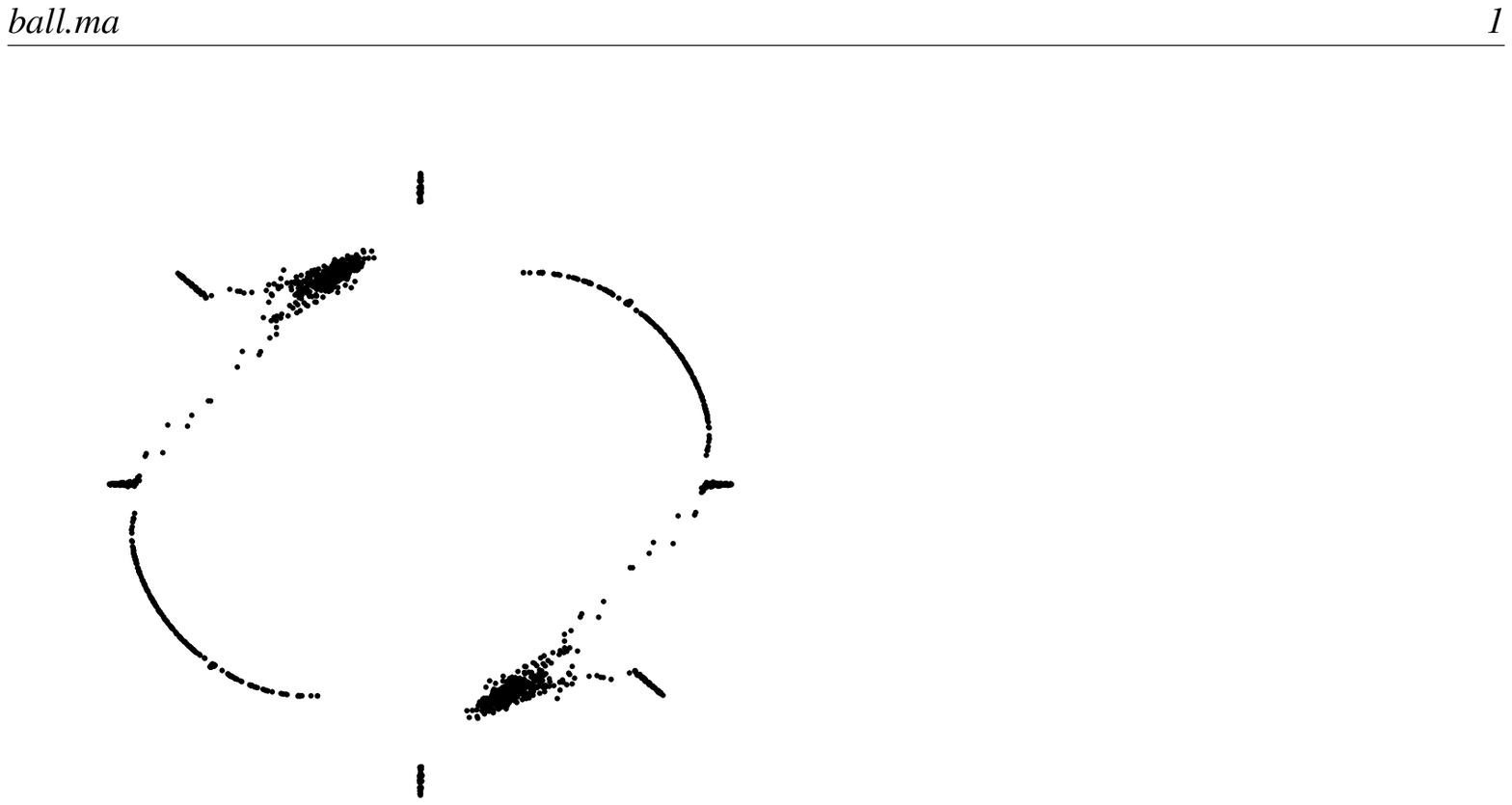,height=5cm,bbllx=90,bblly=510,bburx=300,bbury=710, clip=}
\caption{Distribution of asymptotic velocities $\bar{v}$ for the non--separable 
potential $V(q)=\cos(q_1)+\cos(q_1+q_2)$ and energy $E=3$ (numerical)}
\end{figure}
If for every probability measure $\nu_E$ which is absolutely continuous 
w.r.t.\ $\lambda_E$ 
\[D(\nu_E) := \lim_{t\ar\pm\infty} 
 \frac{\int_\SE \l( q(t,x_0)-q_0 \r)^2 d\nu_E(x_0)} {|t|} \]
exist and is positive, we call $\Phi^t_E$ {\em diffusive}.

Moreover, we say that the flow $\Phi^t_E$ is {\em diffusive in the strong sense}
if $q(t,x_0)/\sqrt{|t|}$ converges weakly to a Gaussian distribution 
with positive covariance matrix. 
An example of a strongly diffusive flow is given in \cite{Kn1}.
\begin{remarks} \label{rema}
\begin{enumerate}
\item \label{one}
Conversely to the third statement, for $d=2$ 
a lowering of the energy may lead to the destruction of the 
second to last 
KAM torus, which in turn may lead to a discontinuous
decrease of the group velocities.

As only for 
$d=2$ the $d$--dimensional  KAM tori have codimension  
one in the $2d-1$--dimensional energy shell $\hSE$,
for $d\geq 3$ Arnold diffusion may lead 
to initial conditions of positive measure which are
not ballistic.
\item 
By the statement 4) there always exists a threshold energy above which the
motion is not diffusive.

\item 
The mean classical velocity $\bar{v}$ for $d=1$ equals the velocity
expectation 
\[\frac{\LA \psi -i\nabla \psi \RA}{\LA \psi,\psi \RA}\]
of the WKB function 
\[\psi(q) := \frac{1}{(E-V(q))^{1/4}} 
\exp\l(\pm i \int_{0}^q \sqrt{2(E-V(q'))} dq'\r).\]
\end{enumerate}
\end{remarks}
\begin{theorem}\label{nonanosov}
If $d\geq2$, then there is no energy $E$ for which 
$\Phi^t_E$ is an Anosov flow.
\end{theorem}
{\bf Proof.} 
If $V=0$, the motion is integrable and thus never Anosov.
So we may assume $V$ non-constant. The Hamiltonian is {\em
optical}, that is, strictly convex on each fibre.

Then for $E\leq \Vmax$ the energy shell $\SE$ 
touches the zero section of $T^*\bR^d$. Thus by Theorem 1 
of Paternain and Paternain \cite{PP} $\Phi^t_E$ is not Anosov.

For $E > \Vmax$ Theorem 3 of \cite{Kn3} which generalizes
a Theorem of E.\ Hopf \cite{Ho} says that the flow $\Phi^t_E$ has
conjugate points if $V$ is non--constant. Thus by Theorem 
1 of \cite{PP} the flow cannot be Anosov either. \hfill $\Box$
\begin{remarks}
\begin{enumerate}
\item 
Of course Theorem \ref{nonanosov} does not imply that motions in
smooth  potentials on $\bT$ cannot be ergodic. To the contrary, 
Donnay and Liverani gave in \cite{DL} a method to construct
such ergodic $C^\infty(\bT)$ potentials for $d=2$ freedoms.
These potentials were constructed in such a way that for a
given energy they contained circularly symmetric pits with
a {\em parabolic} circular trajectory. Any
decrease of the energy then makes this trajectory elliptic and
the motion non-ergodic. 

Our theorem shows that non-hyperbolic trajectories
like in that example must necessarily appear. 
In general we conjecture that for $d=2$ these trajectories
lead to anomalous diffusion effects and are 
incompatible with diffusivity in the strong sense.

Sinai and Kubo gave examples of repelling continuous 
potentials on a torus which lead to ergodic flows.
However, in this case the potentials could not be chosen to be
$C^1$ so that they, too cannot serve as counterexamples to
our theorem. See \cite{DL} for a discussion. 
\item 
Motion of $k$ particles on an $d$--dimensional configuration space
with periodic boundary conditions and mutual forces of
potential type can be described by the motion of one particle 
on a $k\cdot d$--dimensional torus.
Thus Theorem 2 implies that it will be very hard to show
ergodicity of gases if the interparticle forces are smooth.
\item 
A geometric version of the above theorem is:
Geodesic flows on a torus $(\bT,g)$ are never Anosov. 
This follows from the generalizations of Hopf's Theorem
by Burago and Ivanov \cite{BI}, together with Theorem 1 of \cite{PP}.

This is clearly not a mere consequence of the topology 
of the unit tangent bundle $S^{d-1}\times\bT$, 
since for $d=2$ this is a three-torus,
and the simplest example of an Anosov flow (a suspension of Arnold's
cat map) is one on $\bT^3$.  
\end{enumerate}
\end{remarks}
\section{Classical Motion: Coulombic Potentials in $d=2$}
We now treat 
motion in a planar crystal with attracting Coulombic forces.
We fix the locations of the nuclei within the crystal by selecting $m\geq 1$ points $s_1,\ldots,s_m\in \cD$ in
the fundamental domain 
\[\cD:=\{x_1 \ell_1+x_2 \ell_2\mid x_1,x_2\in [0,1)\}\subset \bR^2_\q\]
of the lattice $\cL\subset\bR^2_\q$ with basis $\ell_1,\ell_2$.
The nuclei attract the electron with the {\em charges}
$Z_1,\ldots,Z_m > 0$. That is, we assume the potential of the form
$V(\q)\sim-Z_i/|\q-s_i|$ for $\q$ near $s_i$.
Now by the periodicity of the crystal the potential is singular at
the points of   
\[ \cS := \{ s_i+\ell\mid i \in \{1,\ldots,m\} ,\ell\in\cL \} \]
and thus only defined in the punctured configuration 
plane $\tilde{M} := \bR^2_q\setminus \cS$. Sometimes we identify 
the plane with $\bC$.
\begin{defi}
A potential $V\in C^\infty(\tilde{M},\bR)$ which is $\cL$--periodic
\[V(\q+\ell)=V(\q),\quad (\q\in\tilde{M},\, \ell\in\cL)\]
is called {\bf Coulombic} if for $\vep>0$ small the functions
$f_1,\ldots,f_m$,
\[f_i(Q) :=\l\{
\begin{array}{ll}
V(s_i+Q^2)\cdot Q\bar{Q} &, 0< |Q| <\vep\\
-Z_i       &, Q=0  
\end{array}\right.\]
are $C^\infty$.  
\end{defi}
The reason for the somewhat odd-looking definition is that
we want to regularize the Coulomb singularities by using the
so-called Levi-Civita transformation.  

Observe that $\Vmax := \sup_{\q\in\tilde{M}}V(\q)$ and $\Vmean$
are still well-defined finite quantities.

The classical motion is generated
by the Hamiltonian function
\[\tilde{H}(\p,\q) := \eh \p^2 +V(\q),\quad ((\p,\q)\in T^*\tilde{M}).\]
Due to collisions with the singularities in $\cS$ the Hamiltonian 
flow on the cotangent bundle $T^*\tilde{M}$ of the punctured plane $\tilde{M}$
does not exist for all times. 

However, as described in Lemma \ref{lem:smooth}
below, the flow can be smoothly regularized. 
\begin{lemma} \label{lem:smooth}
There exists a unique smooth extension $(P,\omega,H)$ of the 
Hamiltonian system 
$(T^*\tilde{M}, d\q_1\wedge d\p_1 + d\q_2\wedge d\p_2,\tilde{H})$, where
the phase space $P$ is a smooth four-dimensional manifold with
\beq
P := T^{*}\tilde{M}\cup\bigcup_{\cS} \bR \times S^{1}
\label{Phase:space}
\eeq
as a set, 
$\omega$ is a smooth symplectic two-form on $P$ with
\[\omega\rstr_{T^{*}\tilde{M}} = d\q_1\wedge d\p_1 + d\q_2\wedge d\p_2,\]
and $H:P\ar \bR$ is a smooth Hamiltonian function with
$H\rstr_{T^{*}\tilde{M}} = \tilde{H}$.

The smooth Hamiltonian flow 
\begin{equation}
\Phi^t:P\ar P,\qquad (t\in \bR)
\label{eq:Pt}
\end{equation}
generated by $H$ is complete.

For all energies $E$ which are regular values of $V$, the energy shell 
\begin{equation}
\Sigma_E := \{ x\in P\mid H(x) = E \}
\label{def:SuE}
\end{equation}
is a smooth, three-dimensional manifold, and we write 
$\Phi_E^t := \Phi^t\rstr_{\Sigma_E}$.
\end{lemma}
{\bf Proof.} 
The construction works locally near the singularities $s\in\cS$.
We shortly explain the method by considering the simplest case.
For more details see Prop.\ 2.3 of \cite{KK}, where a scattering potential is
considered.

One may linearize the {\em Kepler flow} with Hamiltonian function 
\[\tilde{H}_K(\p,\q) :=\eh \p^2 - Z/|\q|\] 
in a suitable neighbourhood of $\bR^2_\p\times\{0\}$ in the phase space 
$T^*(\bR^2_\q\setminus\{0\})$, using the 
canonical coordinates $\tilde{H}_K$, $\tilde{T}$, $\tilde{L}$ and 
$\tilde{\varphi}$, where 
$\tilde{L}(\p,\q) := \p\wedge \q$ is the angular momentum, 
$\tilde{T}(\p,\q)$ is the time
needed to come from the phase space point $(\p,\q)$ to the 
pericenter of the Keplerian conic section and $\tilde{\varphi}(\p,\q)$
is the angle between the direction of that pericenter and, say, the
1-direction. Except $\tilde{T}$ these phase space functions are constant along
the Kepler flow, and the collision points correspond to $\tilde{T}=0$,
$\tilde{L}=0$. The remaining coordinates 
$(\tilde{H}_K,\tilde{\varphi})$ take values in a
cylinder $\bR\times S^1$. 

Similarly, because the singularities of $V$ are of the Kepler form,
one may thus complete the phase space
$T^*\tilde{M}$ by gluing one cylinder for each singularity in $\cS$.
\hfill $\Box$\\[2mm]
So after this regularization the Hamiltonian flow exists for all time, 
and we are in a similar situation as in the 
case of a smooth periodic potential treated in Sect.\ \ref{sect:cl:sm}.
In particular we also consider the motion over the configuration torus $\bT$.
\begin{theorem}\label{densevelocities}
If $\Delta \ln(\Eth-V)>0$, then 
for all $E\geq \Eth$ the intersection of the set 
$\bar{v}(\hSE)\subset \bR^2$ 
of asymptotic velocities for energy $E$
with the disk of radius \[\frac{\sqrt{2}(E-\Vmax)}{\sqrt{E-\Vmean}}\]
is dense. 
\end{theorem}
\begin{remark}
Since the motion is diffusive \cite{Kn1} and thus in particular the asymptotic 
velocity can only be non-zero on a set of measure zero,
this should be considered as a `very large deviation' result. 

The ballistic orbits which we construct are of `stop and go' type, that is,
they periodically change between fast motion in a given direction and 
localized motion. Such orbits are somewhat special to our Coulombic potential
and do not exist in the general case. Compare, however, with 
Thm.\ \ref{thm:cl1}.\ref{th:dense}.
\end{remark}
{\bf Proof.}
Although Lemma \ref{lem:smooth} solves the problem of collision orbits and 
regularizes the motion without changing it otherwise, 
we will now introduce a second regularization which is more useful when one tries to construct orbits with prescribed asymptotic velocity, 
whereas it leads to a different phase space and a new time parametrization. 

For Coulombic potentials, 
the Jacobi metric (\ref{Jacobi}) becomes singular near the positions $\cS$
of the nuclei. However this is only a coordinate singularity, and Gaussian  
curvature does not diverge. This can be seen by considering the
case of the Kepler flow and the regularizing complex
coordinate $Q\in\bC$ with $Q^2=q$, 
that is, the {\em Levi-Civita transformation}. It
turns out that in this
coordinate the Jacobi metric, originally defined only for $Q\neq 0$, can be
smoothly extended to a non-singular Riemannian metric
on the whole complex $Q$ plane. Note that the $Q$ plane is a two-fold branched
covering of the original $q$ plane, with a branch point at the position
$q=0$ of the singularity.

In \cite {Kn1} this local construction was globalized
using the toral Riemann surfaces $\bT = M/\cL$, its four-fold covering torus
$\bT^{2} := M/(2\cL)$ (with projection
$\Pi_{\bT^{2},\bT}:\bT^{2}\ar \bT$) 
and the compact Riemann surface 
\beq
\!\!\!\!\!\!\!\!\!\!\!\!\!\! M_4 := \l\{ (q,Q) \in \bT^{2}\times \bP\l| Q^2=
\frac{\prod_{i=1}^m \sigma(q-s_i+\ell_1)\sigma(q-s_i+\ell_1+\ell_2)}
{\prod_{i=1}^m \sigma(q-s_i)\sigma(q-s_i+\ell_2)} \r.\r\},
\label{M:four}
\eeq
\[\sigma(z):= z\prod_{w\in 2\cL\setminus\{0\}} \l(1-z/w\r)\exp\l(
z/w+\eh(z/w)^2\r)\] 
being the Weierstrass $\sigma$--function and $\bP := \bC\cup\infty$ 
the Riemann sphere.
The map $\Pi_{M_4,\bT^{2}}:M_4\ar \bT^{2}$, 
$(q,Q)\mapsto q$ is a two-fold 
branched covering with branch points at the singularities,
all branch numbers equalling one.

Thus by the Riemann-Hurwitz relation the genus $\cG(M_4)$ of $M_4$
equals $\cG(M_4) = 2m+1$ ($m$ being the number of singularities in the
fundamental domain). Since thus the genus is $\geq 3$, by Gauss-Bonnet
the integrated Gaussian curvature 
$\int_{M_4}\KE dM_4= -4\pi\cdot(\cG(M_4)-1)$ 
of the lifted Jacobi metric $g_{4,E}$ on $M_4$
becomes negative. Due to the branched covering construction
the metric $g_{4,E}$, originally not defined at the branch
points, can be {\em smoothly} extended to these points by taking limits.
So the geodesic flow $\phi_{4,E}^t$ on the unit tangent bundle
$T_1M_4$ of the surface $(M_4,g_{4,E})$ is smooth and defined for all times. 

In terms of the potential $V$ the 
Gaussian curvature $\KE$ equals
\begin{equation}
\KE(\q) = \frac{(E-V(\q))\Delta V(\q) + (\nabla V(\q))^{2}}
{2(E-V(\q))^{3}} = -\frac{\Delta \ln(\Eth-V(\q))}{2(E-V(\q))}
\label{Def:K}
\end{equation}
with $\Delta$ and $\nabla$ denoting the Euclidean Laplacian and
gradient, respectively.

For many Coulombic potentials $V$ the Gaussian curvature 
of $M_4$ becomes strictly negative 
($\KE(x)<0$ for all $x\in M_4$) if $E$ is large enough.
Clearly this can only happen if $\Delta V < 0$ in (\ref{Def:K}),
and negativity of (\ref{Def:K}) 
is then preserved if one enlarges the energy $E$.

Similar to the proof of Thm.\ \ref{thm:cl1}.\ref{th:dense}
we construct ballistic orbits in $\bR^2_\q$ by finding closed
geodesics of minimal length. However, due to the presence of 
singularities, these geodesics are not constructed on $\bT$ but on
the smooth Riemannian surface $(M_4,g_{4,E})$.

Since $(M_4,g_{4,E})$ is closed, 
there exists a closed geodesic in every non-trivial 
conjugacy class of the fundamental group $\pi_1(M_4)$. Moreover, 
if the Gaussian curvature $\KE$ is strictly negative
(as it is the case if $\Delta \ln(\Eth-V)>0$ and $E\geq \Eth$), then
this geodesic
$c:S^1\ar M_4$ is essentially unique within its conjugacy class. 
Namely, any other closed geodesic in that class coincides with $c$ up to a
shift of the initial point (see, e.g., Klingenberg \cite{Kl2}, Thm.\ 3.8.14).

As we are interested in ballistic trajectories on $\bR^2_\q$, we seek 
closed trajectories on $\bT$ which are, however, non-contractible 
curves on the configuration torus.
How many such orbits can we construct by projecting closed geodesics
in $M_4$ to $\bT$?

This question can be answered by considering the covering projection
\beq
\Pi_{M_4,\bT} := \Pi_{\bT^{2},\bT}\circ \Pi_{M_4,\bT^{2}}.
\label{def:M4:bT}
\eeq
This continuous map induces a homomorphism 
\[\l(\Pi_{M_4,\bT}\r)_* : \pi_1(M_4)\ar \pi_1(\bT) \]
of fundamental groups. We claim that the image subgroup
${\rm Im}  \subset\pi_1(\bT)$ equals
\beq
{\rm Im} = \pi_1(\bT^{2})\cong 2\cL,
\label{eq:im}
\eeq
that is, consists of the equivalence classes of all loops 
$c:S^1\ar\bT$ which surround the torus $\bT\cong S^1\times S^1$ in both
basic directions an even number of times (this statement is of course
independent of the base). 

(\ref{eq:im}) follows from the definition (\ref{def:M4:bT})
if we show that the homomorphism 
\[\l(\Pi_{M_4,\bT^{2}}\r)_*:\pi_1(M_4)\ar \pi_1(\bT^{2})\] 
is onto. 
To show this, we consider loops in $\bT^{2}$ based at $b\in\bT^{2}$,
where we assume the base point $b$ not to be a one of the $4m$ branch points
of $\Pi_{M_4,\bT^{2}}$, see (\ref{M:four}), so that it has exactly two 
preimage points $b_0,b_1\in M_4$.

Now we can uniquely lift any based loop $c:S^1\ar \bT^{2}$ which 
avoids the branch points,
to a path $\tilde{c}:[0,1]\ar M_4$ starting at $\tilde{c}(0):=b_0$ and
ending either at $\tilde{c}(1)=b_0$ or at $\tilde{c}(1)=b_1$. 

On the other hand, we can find one fixed based loop $l:S^1\ar\bT^2$
which is contractible in $\bT^2$ and is covered by a path 
$\tilde{l}:[0,1]\ar M_4$ connecting the points $\tilde{l}(0) = b_1$
and $\tilde{l}(1) = b_0$. 
Thus either $\tilde{c}$ or $\tilde{c}*\tilde{l}$ ($*$ denoting
concatenation of paths) is a loop in $M_4$ based at $b_0$,
and in both cases the image w.r.t.\ $\Pi_{M_4,\bT^{2}}$ is freely homotopic
to $c$. This shows that the image subgroup ${\rm Im}$ equals
$\l(\Pi_{M_4,\bT^{2}}\r)_*(\pi_1(M_4)) = \pi_1(\bT^{2})$ and thus 
(\ref{eq:im}). 

We need the above information in order to construct fast orbits in a given 
asymptotic direction. But we also need orbits with asymptotic speed
zero in order to construct our `stop and go orbits'. 

To that aim we notice that the kernel
${\rm Ker}(\l(\Pi_{M_4,\bT}\r)_*)\subset \pi_1(M_4)$ of our homomorphism 
is non-trivial. Indeed it contains the commutator subgroup
\[{\rm Comm}(\pi_1(M_4))=\{ghg^{-1}h^{-1} \mid g,h\in \pi_1(M_4)\}\]
of the fundamental group, which is the smallest normal subgroup 
$F$ with $\pi_1(M_4)/F$ abelian. The fundamental group of $M_4$ is non-abelian,
since the genus $\cG(M_4)\geq 3$. So there exists a nontrivial
\[s\in {\rm Ker}(\l(\Pi_{M_4,\bT}\r)_*)\quad,\qquad s\neq{\rm id}.\]
By shortening a loop in the conjugacy class of $s\in\pi_1(M_4)$, we obtain a
closed geodesic $\tilde{s}:S^1\ar M_4$, our {\em stop geodesic}.
After a reparametrization of time the projected geodesic 
$\Pi_{M_4,\bT} \circ \tilde{s}: S^1\ar  \bT$ is a solution curve of the flow 
$\hPhi$ with initial conditions $\hat{x}_s\in\hSE$ 
and period $T_s$. This {\em stop orbit} has asymptotic velocity 
$\bar{v}(\hat{x}_s) = 0$. 

Now in order to show our denseness result, we consider an arbitrary
velocity $v\in\bR^2$ with modulus
\beq
\|v\| \leq \frac{\sqrt{2}(E-\Vmax)}{\sqrt{E-\Vmean}} 
\label{E}
\eeq
and seek for any $\vep>0$ an $\hat{x}_0\in\hSE$ with 
$\| \bar{v}(\hat{x}_0) - v \| < \vep$.

This is easy if $v=0$ because then we set $\hat{x}_0:= \hat{x}_s$
for our stop orbit with $v(\hat{x}_s)=0$.
So we assume that $v\neq 0$ and first approximate the
{\em direction} $v/\|v\|$.

We can find a lattice vector $\ell'\in\cL\cong \pi_1(\bT)$ with
\beq
\l\| \frac{v}{\| v \|} - \frac{\ell'}{\| \ell' \|} \r\| < \frac{\vep}{2}\cdot 
\frac{\sqrt{E-\Vmean}}{\sqrt{2}(E-\Vmax)}.
\label{D}
\eeq
Given $\ell'$, we now construct a closed geodesic $\tilde{g}:S^1\ar M_4$
whose associated periodic orbit on $\bT$ starting at some point 
$\hat{x}_g\in\hSE$ has asymptotic direction
\beq
\frac{\bar{v}(\hat{x}_g)}{\| \bar{v}(\hat{x}_g) \|} = \frac{\ell'}{\| \ell' \|}
\label{B}
\eeq
and speed 
\beq
\| \bar{v}(\hat{x}_g) \| \geq \frac{\sqrt{2}(E-\Vmax)}{\sqrt{E-\Vmean}}.
\label{A}
\eeq
As in the proof of Thm.\ \ref{thm:cl1}.\ref{th:dense} we consider
the closed straight lines 
\[k:S^1\ar\bT,\qquad k(\tau):= q_0+\tau\cdot \ell\, ({\rm mod }\,\cL)\]
with direction $\ell := 4\ell'$ and initial point $q_0\in\bT$. 
Since the loop $k$ is at least four-periodic, we can lift $k$ to 
$M_4$ obtaining a
{\em loop} $\tilde{k}:S^1\ar M_4$ (namely, the lift of $k$ to $\bT^2$
is at least two-periodic and the branched covering $\Pi_{M_4,\bT^{2}}$
is only two-sheeted).

Similar to (\ref{LL}), 
by an  appropriate choice of the initial point $q_0$ we can ensure that
the length 
$L(\tilde{k}) = \|\ell\| \cdot \int_0^1 \sqrt{E-V(k(\tau))}d\tau$
of the corresponding loop in the Jacobi metric is bounded by
\[L(\tilde{k}) \leq \|\ell\|\sqrt{E-\Vmean}.\]

By shortening the loop $\tilde{k}$ we obtain a geodesic
$\tilde{g}:S^1\ar M_4$ which projects to a closed orbit on the torus
starting at some $\hat{x}_g\in\hSE$.

By the argument already used in the proof of Thm.\ \ref{thm:cl1}.\ref{th:dense}
the period $T_g$ of that orbit is 
$\leq \|\ell\| \frac{\sqrt{E-\Vmean}}{\sqrt{2}(E-\Vmax)}$ 
so that the asymptotic 
velocity $\bar{v}(\hat{x}_g) = \ell/T_g$ meets (\ref{A}).
(\ref{B}) is immediate from the construction since 
$\ell'/\|\ell'\|=\ell/\|\ell\|$.

Now this {\em go orbit} is too fast for our purposes.
Therefore we find integers $p,q\in\bN$ with
\beq
\l\| \frac{p}{q} - \frac{T_g}{T_s}\l(\frac{\|\bar{v}(\hat{x}_g)\|}{\|v\|} -1\r)  \r\|
< \delta
\label{p:q}
\eeq
and consider for $n\in\bN$ the group elements 
\[o_n :=  s^{n\cdot p} \cdot g^{n\cdot q}\in \pi_1(M_4).\]
By curve shortening we find a geodesic $\tilde{o}_n:S^1\ar M_4$
in the conjugacy class of $o_n$. We denote the period of the unit speed
reparametrized $\tilde{o}_n$ by $\tilde{T}_n$, and the period 
of the corresponding closed orbit on the torus by $T_n$, and claim
that
\beq
\lim_{n\ar\infty} \frac{T_n}{n(p \cdot T_s+ q\cdot T_g)} = 1.
\label{C}
\eeq
This follows from
\begin{enumerate}
\item
the Anosov property of the geodesic flow on the unit tangent bundle 
$T_1M_4$ of $(M_4,g_{4,E})$ proven in \cite{Kn1} and
\item
the formula $T_n = \int_0^{\tilde{T}_n} \frac{dt}{d\tau} d\tau
=  \int_0^{\tilde{T}_n} \frac{d\tau}{2(E-V(q(\tau)))}$ for the period.
So the time reparametrization factor $1/(2(E-V(q)))$, seen as a 
function on $T_1 M_4$, is H\"{o}lder continuous.  
\end{enumerate}
By 1) the geodesic flow line of $\tilde{o}_n$ approximates
the {\em go geodesic} exponentially in $n$, then switches to the
{\em stop geodesic} in $n$--uniformly bounded time, approximates that
geodesic exponentially in $n$, and finally switches back to the
{\em go geodesic} in $n$--uniformly bounded time.

Thus by 2) the ratio of times in (\ref{C}) goes to one as $n\ar\infty$.

Let $\hat{x}_n\in\hSE$ be a point on the torus orbit corresponding to
the {\em stop and go geodesic} $\tilde{o}_n$. 
Then $\bar{v}(\hat{x}_n) = n q\, \ell/T_n$ and $\bar{v}(\hat{x}_g)=\ell/T_g$
so that by (\ref{C}) 
\[\lim_{n\ar\infty} \|\bar{v}(\hat{x}_n)\| = \frac{\|\ell\|}{T_g+(p/q)T_s}
= \|\bar{v}(\hat{x}_g)\|\l/\l(1+\frac{p T_s}{q T_g} \r)\r. .\] 
The choice (\ref{p:q}) of $p/q$ implies that 
\beq
\lim_{n\ar\infty} \l| \|\bar{v}(\hat{x}_n)\| - \|v\| \r| < \vep/2
\label{F}
\eeq
for $\delta > 0$ small.

The geometric inequality 
\[\|v-\bar{v}(\hat{x}_n)\| \leq | \|\bar{v}(\hat{x}_n)\| - \| v \| |
+ \l\| \frac{v}{\| v \|} - \frac{\bar{v}(\hat{x}_n)}{\| \bar{v}(\hat{x}_n) \|} \r\| \cdot \|v\| \]
together with  (\ref{E}), (\ref{D}), (\ref{B}) and (\ref{F})
gives the result $\|v-\bar{v}(\hat{x}_n)\| <\vep$ for $n$ large. \hfill $\Box$
\section{Semiclassics: Smooth Potentials}
We now compare the quantum system in the semiclassical limit 
with the classical one and thus mimick the definitions of
section 3. 

The Schr\"{o}dinger operators
$H^\hbar(k)$ on $L^2(\bT)$, $k\in\bT^*$ have the eigenvalues
$E^\hbar_n(k)$. The {\em semiclassical asymptotic velocities} are
defined by
\[ \bar{v}^\hbar_n (k) := \l\{ \begin{array}{cl} 
\hbar^{-1}\nabla_k E^\hbar_n(k) & \mbox{gradient exists} \\
0 & \mbox{otherwise.} \end{array}
\r.\] 
We equip the {\em semiclassical phase space} $\hP^\hbar := \bN\times \bT^*$ 
with the {\em semiclassical measure}
$\hl^\hbar := (2\pi\hbar)^{d} \mu_1\times \mu_2$, 
where $\mu_1$ denotes counting measure on $\bN$ and $\mu_2$ Haar measure
on the Brillouin zone $\bT^*$.

In order to compare classical and semiclassical quantities,
we introduce the {\em energy-velocity map} 
\[A^\hbar: \hP^\hbar \ar\bR^{d+1}\quad{\rm with }\quad 
A^\hbar(n,k):=(E^\hbar_n(k),\bar{v}^\hbar_n(k))\]
and the image measure $\nu^\hbar := \hl^\hbar (A^\hbar)^{-1}$.\\[2mm]
Our conjecture, which we shall prove in some special cases is:
\begin{conject} \label{conjecture}
For all $\cL$--periodic potentials $V\in C^\infty(\bR^d,\bR)$
\[w^\ast-\lim_{\hbar\searrow 0} \nu^\hbar = \nu\]
(which means 
\[\lim_{\hbar\searrow 0}\int_{\bR^{d+1}} f(x)d\nu^\hbar(x)=\int_{\bR^{d+1}} f(x)d\nu
(x)\] 
for continuous functions $f\in C^0_0(\bR^{d+1},\bR)$ of compact support).
\end{conject}
As can be already seen from $d=1$--dimensional case, 
the {\em supports} of the semiclassical measures $\nu^\hbar$ are 
in general much larger than the one of $\nu$. If the bands do not touch,
then $\bR\times\{0\}$ belongs to ${\rm supp}(\nu^\hbar)$
(since then by symmetry $\bar{v}^\hbar_n(0)=0$), whereas 
the classical motion is ballistic above $\Vmax$.

In this section we draw conclusions from Birkhoff's Ergodic Theorem 
which for some potentials imply the truth of our conjecture. 

We first show that the range of semiclassical asymptotic 
velocities is included in the {\em convex hull} of the classical ones.
No assumption on the integrability or ergodicity of the classical system is made.

This involves a limit $T\to\infty,\hbar\to0$ which is controlled by the 
Birkhoff type proposition \ref{Birkhoff}.

Let $X$ be a compact metric space, consider a continuous flow
\[\Phi^t: X\ar X \qquad (t\in\bR)\]
and a continuous map 
\[O:X\ar\bR^d.\]

Denote by $M(X)$ the set of Borel probability measures on $X$,
$M(X,\Phi)\subset M(X)$ the set of flow invariant ones and 
\beq
O_T(x) := \frac{1}{T}\int_0^T O\circ\Phi^t(x) dt.
\label{OT}
\eeq
By Birkhoff's Theorem the {\em good set}
\[G:=\{x\in X\mid 
\lim_{T\ar\pm\infty} O_T(x) 
\mbox{ exist and are equal}\, \}\]
has measure $\mu(G)=1$ for all $\mu\in M(X,\Phi)$. We set 
$\bar{O} := \lim_{T\ar\infty} O_T\rstr_G$.

The limit ${\rm dist}(O_T(x),\bar{O}(G))\stackrel{T\to\infty}{\to}0$ is 
in general not uniform in  $x$. However this is true for the convex hull 
\[{\rm conv}(\bar{O}(G)).\] 
Denote for $\cC\subset\bR^d$ and for $\vep>0$ by
$\cC_\vep\subset \bR^d$ the $\vep$--neighbourhood of $\cC$, then it holds:
\begin{propo}\label{Birkhoff}
For all $\vep>0$ there exists  $T_\vep>0$ such that
\[O_T(X)\subset {\rm conv}(\bar{O}(G))_\vep\qquad (|T|>T_\vep).\]  
\end{propo} 
{\bf Proof of \ref{Birkhoff}.}
By compactness of $X$ and thus of $O(X)$ we could otherwise find an $\vep>0$, 
a sequence of points $x_n\in X$ and of times $T_n$ with $T_n\ar\pm\infty$
such that 
\[z := \lim_{n\ar\infty} O_{T_n}(x_n)\in\bR^d\]
exists and $z\not\in \cC_\vep$. W.l.o.g.\ we assume that $T_n\ar +\infty$.

Consider the sequence of probability measures $\mu_n\in M(X)$
given by
\[\mu_n(U):=\frac{1}{T_n} \l|\{ t\in\lbrack0,T_n\rbrack\mid \Phi^t(x_n)\in U  \}\r|
\qquad (U\subset X\,{\rm Borel}).\] 
We now use the following facts (see Thm.\ 6.10 in 
Walters' book, \cite{Wa}):
$M(X)$ and $M(X,\Phi)$ are  non-empty, convex, and compact in the
weak--*--topology. The extreme
points of $M(X,\Phi)$ coincide with the ergodic measures.
By going to a subsequence, if necessary, 
\[\mu := w^\ast-\lim_{n\ar\infty} \mu_n\in M(X)\]
exists by compactness of $M(X)$. $\mu\in M(X,\Phi)$ and as 
$O_{T_n}(x_n)=\int O d\mu_n$
the expectation $\int_X O d\mu=z\not\in \cC_\vep$. By Choquet decomposition 
of $\mu$ we would find an ergodic measure $\nu\in M(X,\Phi)$ with 
 $\int_X O d\nu\not\in \cC_\vep$. On the other hand by ergodicity of
$\nu$ there exists an $x\in X$ with $\bar{O}(x)=\int_X O d\nu$, which is a 
contradiction. \hfill $\Box$\\[2mm]
For $I\subset \bR$ compact the phase space region
\[\hP_I := \{ x\in\hP \mid H(x)\in I\}\]
is compact and $\hPhi$--invariant so that we are in the situation of Prop.\
\ref{Birkhoff}.

The semiclassical analog of the thickened energy shell $\hP_I$
is 
\[\hP^\hbar_I:= \{(m,k)\in \hP^\hbar\mid E_m^\hbar(k)\in I\}.\]
We equip
them with the probability measures 
\[\hl_I := \frac{\hl}{\hl(\hP_I)}\quad\mbox{on}\quad \hP_I\]
and (for $\hbar$ small)
\[ \hl^\hbar_I := \frac{\hl^\hbar}{\hl^\hbar (\hP^\hbar_I)}
\quad\mbox{on}\quad \hP^\hbar_I.\]
These induce the image probability measures $\mu_I := \hl_I \bv^{-1}$ and
$\mu^\hbar_I := \hl^\hbar_I (\bv^\hbar)^{-1}$ on the space $\bR^d$ of asymptotic
velocities.

We shall now consider for $\vep>0$ intervals
\[I_\vep := \lbrack E-\vep, E+\vep\rbrack.\]
and show our semiclassical results on the group velocities: 
\begin{theorem} \label{semi:two}
Let $V\in C^\infty(\bR^d, \bR)$ be $\cL$--periodic, $E\in\bR, \vep>0$.
\begin{enumerate}
\item
Let $\cC:={\rm conv}(\bar v(\hat P_{I_{2\vep}}))$. Then for all $\eta>0$, a.e. $k\in
\bT^\ast$
$\exists\hbar_0\ \forall\hbar\le\hbar_0$ \[\bar{v}_j^\hbar\in \cC_\eta\qquad {\rm if\
} E_j^\hbar(k)\in I_\vep\ ;\]  %
\item 
let $\cS := {\rm
conv}({\rm supp}(\mu_{I_{2\vep}}))\subset\bR^d$ be the convex hull of the support of
$\mu_{I_{2\vep}}$,  then the semiclassical measures concentrate inside $\cS$:
For all $\eta>0$
\[\lim_{\hbar\searrow 0} \mu^\hbar_{I_\vep} (\cS_\eta) = 1.\]
\end{enumerate}

\end{theorem}
\begin{remark}
In general $\cS\subset \overline{\cC}$ is much 
smaller than 
$\overline{\cC}$. As an example for ergodic motion one has by 
Thm.\ \ref{thm:cl1}.\ref{th:erg}
$\cS=\{0\}$, whereas by Thm.\ \ref{thm:cl1}.\ref{th:dense} 
$\overline{\cC}$ contains a disk of radius 
$\frac{\sqrt{2}(E-\Vmax)} {\sqrt{E-\Vmean}}$.
\end{remark}
{\bf Proof. }
The proof of 1.\ is based on Theorem \ref{Birkhoff}, whereas for 2.\
we use the a.e convergence to the asymptotic velocity and a Shnirelman
type argument. We shall freely use the semiclassical calculus as exposed in
\cite{robe},  \cite{helfrobemart} and references therein. 

First we state a lemma about Bloch decomposition of Anti-Wick quantization.
\begin{lemma}\label{antiwick}
Let $f\in C_b^\infty(\bR^{2d},\bR),\quad f(p,q+\ell)=f(p,q)\qquad(\ell\in\cL)$,
\[f^{AW}\psi:=\int_{\bR^{2d}}f(p,q)\phi_{p,q}
\langle\phi_{p,q} ,\psi\rangle\ 
{dp\, dq\over{(2\pi\hbar)}^d}\qquad(\psi\in L^2(\bR^d))\]
\[\hbox{where\ }\phi_{p,q}(x):=e^{-{i\over2\hbar}pq}e^{{i\over\hbar}px}\phi(x-q), 
\quad \phi(x):=(\pi\hbar)^{-d/4}e^{-{x^2\over2\hbar}}.\]
It holds
\[Uf^{AW}U^{-1}=\int^\oplus_{\bT^{*}} f^{AW}(k){dk\over\vert\bT^{*}\vert}\]
with
\[f^{AW}(k)\psi:=\int_{\hP} f(p,q) U\phi_{p,q}(k)
\langle U \phi_{p,q}(k), 
\psi\rangle_{L^2(\bT)}{dp\, dq\over{(2\pi\hbar)}^d}\]
\end{lemma}
{\bf Proof.}
By periodicity of $f$ and unitarity of $U$ we have 
\[Uf^{AW}\psi=\sum_{\ell\in\cL}\int_{\hP}f(p,q)
U\phi_{p,q+\ell}\langle U\phi_{p,q+\ell},U\psi\rangle{dp\, dq\over{(2\pi\hbar)}^d}.\]
Now
\[U\phi_{p,q+\ell}(k,x)=e^{i(p+\hbar k)\ell/\hbar}e^{-i/(2\hbar)\ell
p}U\phi_{p,q}(k,x)\] 
so
\[Uf^{AW}\psi(k)=\]
\[\hspace{-5mm} \int_{\hP}f(p,q)U\phi_{p,q}(k)\sum_{\ell\in \cL}
\int_{\bT^\ast}{dk'\over\vert\bT^\ast\vert}
(e^{i(k-k')\ell}\langle U\phi_{p,q}(k'),U\psi(k')\rangle_{L^2(\bT)})
{dp\, dq\over{(2\pi\hbar)}^d}\]
the claim follows now from Fourier inversion and the 
$\cL^\ast$--periodicity of\\ 
$k'\mapsto\langle U\phi_{p,q}(k'),U\psi(k')\rangle$. \hfill $\Box$\\[5mm]
A corollary of this lemma, the Egorov Theorem and the 
Weyl--Anti-Wick correspondence
is:
\[e^{iH^\hbar(k)t/\hbar}f^{AW}(k)e^{-iH^\hbar(k)t/\hbar}=
(\hat f\circ\hat\Phi^t)^{AW}(k)+ {\cal O}_T(\hbar).\]

Denote $\Lambda^\hbar_\vep(k)=\lbrace j; E_j^\hbar(k)\in I_\vep\rbrace$ and
$\chi\in C_0^\infty(\bR,\bR)$, ${\rm supp\ }\chi\subset I_{2\vep}$, 
$\chi\rstr_{I_\vep}=1$.
For $j\in \Lambda^\hbar_\vep(k)$ consider the eigenfunction $\psi_{j,k}^\hbar$
of $H^\hbar(k)$ and its Husimi distribution 
$\rho_{j,k}^\hbar: P\ar\bR$, 
\[\rho_{j,k}^\hbar(p,q) := (2\pi\hbar)^{-d}
\vert\langle U\phi_{p,q}(k),\psi_{j,k}^\hbar\rangle_{L^2(\bT)}\vert^2.\]

We then have (using notation (\ref{OT}))
\[\bar v_j^\hbar(k)=\int_{\hP}(\chi(H)p)_T \, d\rho_{j,k}^\hbar\,+\,
{\cal O}_T(\hbar).\]

We apply Prop.\ \ref{Birkhoff} with $X=\hat P_{I_{2\vep}}$. By time reversal
symmetry $0\in {\rm conv}(\bar v(X))$ so we find a $T$ such that
\[(\chi(H)p)_T(X)\subset {\rm conv}(\bar v(X))_{\eta/2}\]
and an $\hbar$ such that $O_T(\hbar)<\eta/2$. Thus 1. is proven.

Now we show that for a.e.\ $k$, for all $\eta>0$
\beq
\lim_{\hbar\searrow 0} 
\frac{\#\l\{m\in\Lambda^\hbar_\vep(k);
\l|\bv^\hbar_m(k)-\int\chi(H)\bar v d\rho_{m,k}^\hbar\r|<\eta\r\}}
     {\# \Lambda^\hbar_\vep(k)} = 1
\label{LaLa}
\eeq

This implies 2.\ as $\int\chi(H)\bar v d\rho_{j,k}^\hbar\subset{\rm 
conv}({\rm supp}\mu_{I_{2\vep}})$ by absolute continuity of 
$\chi(H)d\rho_{j,k}^\hbar$ w.r.t.
$\hat\lambda_{I_{2\vep}}$, by time reversal symmetry and the bound $1$ for the
fraction.

By Birkhoff's Theorem 
\[\lim_{T\ar\infty}
\int_\hP \l|\hat{v}_T(\hx) -\bv(\hx) \r|
d\hl_{I_{2\vep}} =0.\]
So the set 
\[B(T,\eta) := 
\l\{ \hx\in\hP_{I_{2\vep}}\l|\ \l|\hat{v}_T(\hx) -\bv(\hx) \r| \geq
\frac{\eta}{4}\r.
\r\}\]
of phase space points $\hx$ eventually giving rise to 
$\hat{v}_T(\hx)\not\in \cS_{\eta/2}$
can be made small: 
\[\hl_{2\vep}(B(T,\eta)) \leq \frac{\delta}{2}\qquad (T\geq T(\delta)). \]
On the other hand, by  convergence on the $\hbar$--independent
$\hat P_{I_{\vep}}$ : 
\[w^\ast-\lim_{\hbar\searrow 0}\ \frac{1}{\#\Lambda^\hbar_\vep(k)}\!\!
\sum_{m\in\Lambda^\hbar_\vep(k)} \rho_{m,k}^\hbar = \hl_{I_{\vep}}.\] 
For $\hbar\leq\hbar(\delta)$ and $T\geq T(\delta)$ we thus have
\[ \frac{1}{\# \Lambda^\hbar_\vep(k)}
\sum_{m\in\Lambda^\hbar_\vep(k)} \rho_{m,k}^\hbar\, (B(T,\eta)) \leq \delta. \]
By Tchebycheff's inequality
\[\# \l\{ m\in\Lambda^\hbar_\vep(k)\mid  \rho_{m,k}^\hbar\, (B(T,\eta)) 
\geq\sqrt{\delta}  \r\}  \leq \sqrt{\delta}\  \#\Lambda^\hbar_\vep(k)\ .\]
For $m$ in the complementary set it holds:
\[\int_{\hP}\chi(H)(p_T-\bar v) d\rho_{m,k}^\hbar\le 
{\eta\over4}+2\Vert\chi(H)\bar
v\Vert_\infty \sqrt{\delta}<{\eta\over2}\qquad(\delta<\delta(\eta))).\]

To summarize: for $\alpha>0$ there is a $T$ and a set
$G_T^\hbar\subset\Lambda^\hbar_\vep$ 
such that for $\hbar$ small enough $\vert\bar
v_j^\hbar-\int \chi(H)\bar v d\rho_{j,k}^\hbar\vert<\eta$  for $j\in G^\hbar_T$ and 
${\#G^\hbar_T\over\#\Lambda^\hbar_\vep}\ge 1-\alpha$. 
This finishes the proof of 2.
\hfill $\Box$
\begin{coro}
If the classical motion is non-ballistic with probability one on an energy 
interval $I$: $\mu_I=\delta_0$,
then Conjecture \ref{conjecture} holds true:
\[w^\ast-\lim_{\hbar\searrow 0} \nu^\hbar = \nu.\]
\end{coro}
For example, this is the case if the classical motion is ergodic.
\section{Semiclassics: Separable Potentials}
If the potential is separable, the distribution of semiclassical 
group velocities converges rapidly to the classical velocity distribution.
We begin with the case of one dimension, and  
thus consider the operator $H := -\hbar^2 \frac{d^2}{dx^2} + V(x)$
with potential $V\in C^r(\bR,\bR)$, assuming w.l.o.g.\ that $V(x+1)=V(x)$.
The band function of the $n$--th band for the quasimomentum 
$k\in[-\pi,\pi]$ is denoted by $E_{n}(k)\equiv E^\hbar_n(k)$. 
Of course $\frac{dE_n}{dk}(k)=0$ or the band functions touch at the band edges
$k=0$ or $\pm \pi$. However, apart from small neighbourhoods
of these values of the quasimomentum it holds: 
\begin{propo}\label{d1semiclassics}
Assume that the periodic potential $V\in C^r(\bR,\bR)$, $r\geq 2$.
\begin{enumerate}
\item
Then all bands in the energy interval $[\Vmax+\vep,\infty)$ meet the 
following uniform estimate. If the quasimomentum 
$|k|\in [\hbar^{(r-1)/2},\pi-\hbar^{(r-1)/2}]$, then 
\[\l| {\rm sign}(k)\cdot\hbar^{-1}\frac{dE_n}{dk}(k) - 
(-1)^n v_{\rm cl}(E_n(k)) \r| \leq c\hbar\]
for some $c=c(\vep)>0$,
\[v_{\rm cl}(E) = 
\eh \l(\int_0^1\frac{1}{\sqrt{E-V(t)}}dt\r)^{-1}\qquad (E>\Vmax),\]
$v_{\rm cl}(E)=0$ $(E\leq\Vmax)$ being the absolute value of the classical velocity.
\item
In the energy range $[\Vmin,\Vmax-\vep]$ and for all $k\in[-\pi,\pi]$
\[ \frac{dE_n}{dk}(k) = \OO{\infty}. \]
\item
\beq
w^\ast-\lim_{\hbar\searrow 0} \nu^\hbar = \nu. 
\label{nu:sep}
\eeq
\end{enumerate}
\end{propo}
{\bf Proof.}\\ 
{\bf 1)}
We first consider the energy interval $[\Vmax+\vep,\infty)$.
For the $k$ values under consideration, the Bloch eigenfunctions have
no zeroes. So we are looking for zero--free solutions
$\ph:\bR\ar\bC$ of the differential equation 
\beq
H\ph = E \ph.
\label{HphEph}
\eeq
The complex phase $S:=\frac{\hbar}{i}\ln(\ph)$  of such a solution
solves the differential equation 
\beq
(S')^2 - i\hbar S'' - W = 0
\label{S:Dgl}
\eeq
with $W := E-V$. 
We solve this equation, using the Ansatz
\[S(x)=\tilde{S}_r(x)+ \hbar^{r+1}R(x,\hbar)\quad  \mbox{with} \quad 
\tilde{S}_r(x) := \sum_{n=0}^r \hbar^n S_n(x).  \]
With
\[S_0(x) := \int_0^x \sqrt{W(t)}dt \]
the recursion equation
\beq
iS_{n-1}'' - \sum_{l=0}^n S_l'S_{n-l}' = 0\qquad (n=1,\ldots,r) 
\label{rekursion}
\eeq 
has the continuous solution 
\[S_n(x) = 
\eh \int_0^x 
\frac{iS_{n-1}''(t)-\sum_{l=1}^{n-1} S_l'(t)S_{n-l}'(t)} {\sqrt{W(t)}} dt 
\qquad (n=1,\ldots,r),\]
since $S_0'=\sqrt{W}>0$.

In particular we have $S_1(x) = i\ln(\sqrt[4]{W(x)})+c$. We set $c:=0$.
Then on bounded intervals 
\[\ph(x)= \frac{1}{\sqrt[4]{W(x)}} \exp \l(i \int_0^x \sqrt{W(t)} dt/\hbar \r) + \OO{1}.\]
As a consequence of (\ref{rekursion}) $S_n$ is real if $n$ is even 
and imaginary if $N$ is odd.
$\tilde{S}'_r$ is 1--periodic.

But unlike the real part of $\tilde{S}_r$ 
the imaginary part is always 1-periodic.
This can be seen, e.g., by considering the formal power series
$\tilde{S}_\infty$ in $\hbar$. By (\ref{S:Dgl})
the real part
$R := \Re(\tilde{S}'_\infty)$ of the derivative is 
related to the imaginary part
$I := \Im(\tilde{S}'_\infty)$ by
\[\hbar R' = 2IR,\]
so that $I=\eh \hbar (\ln(R))'$. 
Thus the formal power series
$\Re(\tilde{S}_\infty)$
is 1-periodic, 
which implies the same periodicity for its coefficients
$S_{2n+1}$.

Of course this argument is even valid if (by finite differentiability
of $V$) only finitely many coefficients
$S_{n}$ are defined.

Thus for $E > \Vmax$ 
\[\tilde{\ph}_r :=\exp\l(\frac{i}{\hbar}\tilde{S}_r\r)
= A_r\exp\l(\frac{i}{\hbar}U_r\r):\bR\ar\bC\]
is a function with periodic modulus $A_r>0$ and phase
$U_r = \Re(\tilde{S}_r)$. 
We compare this with the solution 
$\ph$ of (\ref{HphEph}) with the initial values
\[\ph(0):=\tilde{\ph}_r(0),\quad \ph'(0):=\tilde{\ph}'_r(0).\]
On bounded spatial intervals one has the estimate uniform in $x$ and $E$
\[ \ph(x) = \tilde{\ph}_r(x)+\OO{r},\quad 
   \ph'(x) = \tilde{\ph}'_r(x)+\OO{r}.\]  
The same is true for the matrix
\[ M(x) := 
\l(\begin{array}{cc} \ph(x) & \bar{\ph}(x)\\ \ph'(x) & \bar{\ph}'(x)
\end{array}\r) \] 
of the corresponding fundamental system ($\bar{\ph}$ is linearly independent
of $\ph$). The monodromy matrix $T := M(1) \cdot M(0)^{-1}$ has
determinant 1 and trace
\[{\rm Tr}(T) = 2\Re(\tilde{\ph}_r(1)/\tilde{\ph}_r(0)) + \OO{r}=
2\cos(U_r(1)/\hbar) + \OO{r}.\]
Now we consider those quasimomenta $k$ for which 
\[{\rm dist}\l(U_r(1)/\hbar,\pi\cdot\bZ \r) > \hbar^{(r-1)/2}.\]
For them 
$|{\rm Tr}(T)| < 2$ if $\hbar< \hbar_0$ 
so that the quasiperiodic function $x\mapsto \ph(x)$ is bounded.
Thus $\ph$ is a Bloch function with quasimomentum $k$, 
\[\cos(k)=\eh{\rm Tr}(T).\] 
Differentiating both sides w.r.t.\  $k$ yields
\beq
\sin(k) =  \sin(U_r(1)/\hbar)\cdot \frac{dU_r(1)}{dE}\cdot
\hbar^{-1}\frac{dE}{dk} +\OO{r}.
\label{alt}
\eeq
On the other hand 
\[\sin(k) = \sqrt{1-(\eh{\rm Tr}(T))^2} = 
\pm \sin(U_r(1)/\hbar)\cdot(1+\OO{1}),\]
so that (\ref{alt}) implies the relation
\beq
\hbar^{-1}\frac{dE}{dk} = \pm \l(\frac{dU_r(1)}{dE}\r)^{-1}
+\OO{1}
\label{neu}
\eeq
for the group velocity. 
By
$U_r(1) = \int_0^1 \sqrt{W(t)}dt + \OO{1}$
we get the estimate
\[\hbar^{-1}\frac{dE}{dk} = 
\pm \eh \l(\int_0^1\frac{1}{\sqrt{E-V(t)}}dt\r)^{-1} 
+\OO{1}\]
for quasimomenta $|k|\in[\hbar^{(r-1)/2},\pi-\hbar^{(r-1)/2}]$. 
\\[5mm]
{\bf 2)} For $E\leq\Vmax-\vep$ we have $v_{\rm cl}(E) = 0$. 
Quantum mechanically it is well-known that the wave function 
and its derivatives are exponentially
decreasing w.r.t.\ $\hbar$ well inside the potential well, say, for $V(x)\leq \vep/2$.
This implies exponential decay of the group velocity. 

Namely
\[\hbar^{-1}\frac{dE_n}{dk}(k)= \int_0^1 j_n(x)dx = j_n(x)\]
for the current $j_n(x)= -i\hbar\Im(\bar{\ph}_n(k)(x)\frac{d}{dx}\ph_n(k)(x))$ of the
eigenfunction $\ph_n(k)$ with eigenvalue $E_n(k)\leq\Vmax-\vep$,
since the divergence of the current of eigenfunction vanishes.

If we evaluate $j$ at $x$ inside the potential well, then we see that it is
exponentially small. For more precise estimates valid in the multidimensional case
we refer to Outassourt \cite{Ou}. 
\\[5mm] 
{\bf 3)}
The convergence of the semiclassical measures $\nu^\hbar$ to $\nu$
follows from the following reasoning.

By a Weyl estimate we have weak--*--convergence in energy distribution:
\[w^\ast-\lim_{\hbar\searrow 0} \hl^\hbar (E^\hbar)^{-1} = \hl H^{-1}\]
with $E^\hbar:\hP^\hbar\ar\bR$, $E^\hbar(m,k) := E^\hbar_m(k)$ being the energy function
on the Fermi surface. 
The first two parts of the proposition exclude the energy interval
$(\Vmax-\vep,\Vmax+\vep)$. However, as $\hbar\searrow 0$,
we can let $\vep\equiv\vep(\hbar)\searrow 0$, too. By the above Weyl estimate
we do not loose anything of $\nu^\hbar$ in the semiclassical limit
\[ \lim_{\hbar\searrow 0} \sum_{m\in\bN} 
(2\pi\hbar)^{-1}\l|\{ k\in\bT^* \mid E_m^\hbar(k)-\Vmax| < \vep(\hbar)\} \r| =0.\]
Then (\ref{nu:sep}) follows from {\bf 1)} and {\bf 2)}. 
\hfill $\Box$
\begin{coro}
Let $V\in C^2(\bR^d,\bR)$ be a separable periodic potential.
Then Conjecture \ref{conjecture} holds true:
\[w^\ast-\lim_{\hbar\searrow 0} \nu^\hbar = \nu. \]
\end{coro}
{\bf Proof.} 
By our assumption the potential is of the form 
\[V(\q) = \sum_{j=1}^d V_j(q_j)\]
with $V_j\in C^2(\bR,\bR)$ of some period $l_j>0$.
Let $\nu^\hbar_j$ and $\nu_j$ denote the (semi)classical measures
for the one-dimensional potential $V_j$.
Then 
\[ \nu = (\nu_1,\ldots,\nu_d) L^{-1}\quad\mbox{and}\quad
\nu^\hbar = (\nu^\hbar_1,\ldots,\nu^\hbar_d) L^{-1}\]
for the linear map 
$L:\l(\bR^2\r)^d\ar\bR^{d+1}$, 
\[L(h_1,v_1,h_2,v_2,\ldots,h_d,v_d)\mapsto (h_1+\ldots+h_d,v_1,\ldots,v_d). \]
Although for $d>1$ the linear map $L$ is not injective and
thus not proper, its restriction to
\beq
\times_{j=1}^d \l( [V_{j,{\rm min}},\infty)\times \bR \r)
\label{last}
\eeq
has this property, so that preimages of compactly supported functions are 
still compactly supported. We may restrict $L$ to (\ref{last}), since
the support of $(\nu_1,\ldots,\nu_d)$ is contained in (\ref{last}),
the spectrum of $H_j$ being contained in $[V_{j,{\rm min}},\infty)$

So multidimensional convergence follows from the one-dimensional one.
\hfill $\Box$

\end{document}